\documentclass[12pt]{article}

\usepackage{epsfig,multicol,multirow,cite}
\usepackage{amsmath,subfigure,latexsym,amssymb}
\usepackage[table]{xcolor}

\newcommand{\be}{\begin{equation}}
\newcommand{\ee}{\end{equation}}
\newcommand{\nn}{\nonumber}
\newcommand{\bea}{\begin{eqnarray}}
\newcommand{\eea}{\end{eqnarray}} 

\newcommand{\la}{\langle}
\newcommand{\ra}{\rangle}
\newcommand{\uno}{1 \!\! 1}
\newcommand{\Z}{\mathbb{Z}}
\newcommand{\R}{{\kern+.25em\sf{R}\kern-.78em\sf{I} \kern+.78em\kern-.25em}}
\newcommand{\RR}{{\kern+.25em\sf{R}\kern-.6em\sf{I} \kern+.6em\kern-.25em}}
\newcommand{\N}{{\kern+.25em\sf{N}\kern-.78em\sf{I} \kern+.78em\kern-.25em}}
\newcommand{\C}{{\kern+.25em\sf{C}\kern-.50em\sf{I} \kern+.50em\kern-.25em}}

\newcommand{\ri}{{\rm i}}

\newcommand{\vp}{\varphi}

\newcommand{\ltapprox}{\raisebox{-0.5ex}{$\,\stackrel{<}{\scriptstyle\sim}\,$}}

\usepackage{color}
\definecolor{col_red}{rgb}{1.0,0.0,0.0}

\makeatletter
\@addtoreset{equation}{section}
\makeatother

\begin{document}
 
\begin{center}
{\Large\bf Interpreting Numerical Measurements}

\vspace*{6mm}

{\Large\bf in Fixed Topological Sectors} \\

\vspace*{1cm}

Wolfgang Bietenholz$^{\rm a}$, Christopher Czaban$^{\rm b}$,
Arthur Dromard$^{\rm b}$, \vspace*{1mm} \\ Urs Gerber$^{\rm a,c}$, 
Christoph P.\ Hofmann$^{\rm d}$, \vspace*{1mm} \\
H\'{e}ctor Mej\'{\i}a-D\'{\i}az$^{\rm a}$ and Marc Wagner$^{\rm b}$ \\
\ \\
$^{\rm \, a}$  Instituto de Ciencias Nucleares \\
Universidad Nacional Aut\'{o}noma de M\'{e}xico \\
A.P. 70-543, C.P. 04510 Ciudad de M\'{e}xico, Mexico\\
\ \\ \vspace{-3mm}

$^{\rm \, b}$ Goethe-Universit\"{a}t Frankfurt am Main\\
Institut f\"{u}r Theoretische Physik \\
Max-von-Laue-Stra\ss e 1, D-60438 Frankfurt am Main, Germany\\
\ \\ \vspace{-3mm}

$^{\rm \, c}$ Instituto de F\'{\i}sica y Matem\'{a}ticas \\
Universidad Michoacana de San Nicol\'{a}s de Hidalgo\\
~~~~Edificio C-3, Apdo.\ Postal 2-82, C.P.\ 58040, 
Morelia, Michoac\'{a}n, Mexico \\ \ \\ \vspace{-3mm}
 
$^{\rm \, d}$ Facultad de Ciencias, Universidad de Colima\\
Bernal D\'{\i}az del Castillo 340, Colima C.P. 28045, Mexico

\end{center}

\vspace*{6mm}

\noindent
For quantum field theories with topological sectors, Monte Carlo 
simulations on fine lattices tend to be obstructed by an
extremely long auto-correlation time with 
respect to the topological charge. Then reliable numerical 
measurements are feasible only within individual sectors. 
The challenge is to assemble such restricted measurements in a way
that leads to a substantiated approximation to the fully fledged result, 
which would correspond to the correct sampling over the entire set of 
configurations. We test an approach for such a topological summation, 
which was suggested by Brower, Chandrasekharan, Negele and Wiese. 
Under suitable conditions, energy levels and susceptibilities can be
obtained to a good accuracy, as we demonstrate for O($N$) models,
SU(2) Yang-Mills theory, and for the Schwinger model.

\newpage

\tableofcontents

\section{Motivation}

We consider quantum field theories with topological sectors, 
in Euclidean spacetime. These sectors are characterized by a topological 
charge $Q \in \Z$, which is a functional of the field configuration.
In infinite volume, the configurations with finite
action are divided into these disjoint sectors. The same property
holds in finite volume with periodic boundary conditions.

Examples are O$(N)$ models in $d = N-1$ dimensions, all 2d CP$(N-1)$ 
models, 4d SU$(N)$ Yang-Mills gauge theories ($N\geq 2$), as well as 
QCD, and 2d U(1) gauge theory, as well as the Schwinger model.
In all these models, a continuous deformation of 
a given configuration (at finite action)
can only lead to configurations within the same topological sector, 
{\it i.e.}\ the deformation cannot alter the topological charge $Q$.

In light of this definition, lattice regularized models have
in general no topological sectors --- strictly speaking. 
Nevertheless, it is often useful to 
divide the set of lattice field configurations into sectors, which turn
into the topological sectors in the continuum limit. The definition
of a topological charge on the lattice is somewhat arbitrary.
In presence of chiral fermions (where the lattice Dirac operator
obeys the Ginsparg-Wilson relation), the fermion index provides
a sound formulation \cite{Has98}.
For the O$(N)$ models the geometric definition \cite{BergLuscher} 
is optimal, since it guarantees integer topological charges on periodic 
lattices (for all configurations except for a subset of measure zero).
In gauge theory, field theoretic definitions are often applied,
usually combined with smearing or cooling techniques,
see {\it e.g.}\ Ref.\ \cite{Forcrand97}. These techniques are
computationally cheap and provide, on fine lattices or at fixed 
topology, results which agree well with the computationally demanding
fermion index \cite{Bruck10,Cichy14,CichyLat14}.

As we proceed to finer and finer lattices, the formulation
becomes more continuum-like, and changing a (suitably defined)
topological sector of the lattice field is getting more and more
tedious --- for this purpose, continuous deformations have to 
pass through a statistically suppressed domain of high Euclidean 
action. To a large extent, this property persists for finite but 
small deformations, as they are carried out in the Markov chain
of a Monte Carlo simulation which performs small update steps.

In QCD simulations with dynamical quarks, the gauge configurations
are usually generated with a Hybrid Monte Carlo (HMC) algorithm, 
with small updates, on lattices of a spacing $a$ in the range 
$0.05~{\rm fm} \ltapprox a \ltapprox 0.15~{\rm fm}$.
The artifacts due to the finite lattice spacing tend to be
the main source of systematic errors. Therefore, the lattice 
community will try to suppress them further by 
proceeding to even finer lattices, $a < 0.05~{\rm fm}$. 

This will provide continuum-like features, which are highly
welcome in general, but as a draw-back it will become
harder to change the topological sector. A HMC simulation may 
well be trapped in a single sector over a tremendously long 
trajectory; in particular, this is the experience in QCD 
simulations with dynamical overlap quarks \cite{JLQCD}. 
In this case, Ref.\ \cite{Egri} suggested a method to estimate
the ratio between topologically constrained partition functions,
and tested this method by determining the topological
susceptibility from fixed topology overlap quark simulations.

In some circumstances it is even motivated to suppress topological 
transitions on purpose, in particular when dealing with dynamical 
chiral fermions. In that context,
configurations in a transition region cause technical problems,
like a bad condition number of an overlap or domain wall Dirac operator. 
This can be avoided by the use of unconventional lattice gauge actions, 
known as ``topology conserving gauge actions'' \cite{topogauge,Bruck10}
(see also Ref.\ \cite{Holland} for a very similar formulation).

A further option is the use of a ``mixed action'', where one implements 
chiral symmetry only for the valence quarks, which requires just
a moderate computational effort. In particular, overlap valence quarks
have been combined with Wilson sea quarks. However, in this set-up
the continuum limit is not on safe ground, because (approximate)
valence quark zero modes are not compensated by the sea quark spectrum 
\cite{Cichy10}. This problem might be avoided by fixing the 
topological sector particularly to $Q=0$.

In such settings, there are obvious questions about the (effective) 
ergodicity of the algorithm, since the simulation does not sample 
properly the entire space of all configurations. Even if we ignore 
this conceptual question, in practice the measurement of an 
observable may well be distorted. This is the issue to be addressed 
in this work.

Section 2 describes the Brower-Chandrasekharan-Negele-Wiese (BCNW) 
approach, and Sections 3  and 4 probe  it in the 1d O(2) and the 
2d O(3) non-linear $\sigma$-model. 
It is explored further in 4d SU(2) Yang-Mills theory in Section 5,
and in the Schwinger model in Section 6. The field theoretic models
discussed in Sections 4 to 6 share fundamental features with QCD. 
Section 7 is devoted to our conclusions.

\section{The BCNW method}

As a remedy against the topological freezing of Monte Carlo histories, 
L\"{u}scher suggested the use of open boundary conditions, such that
the topological charge can change continuously \cite{openbc}. 
This overcomes the problem, but it breaks translational 
invariance\footnote{A recent work \cite{Mages} suggests the use of 
P-periodic (instead of open) boundary conditions in Euclidean time
{\it i.e.}\ a parity flip, which also implies $Q \in \R$, but 
translation symmetry breaking effects are exponentially suppressed.}
and one gives up integer topological charges $Q$. 
However, $Q \in \Z$ provides a valuable
link to aspects, which are analytically known or conjectured
in the continuum, for instance regarding the $\epsilon$-regime of QCD, 
or properties based on an instanton picture.
Therefore it is still motivated to also explore alternative approaches.

In this work we maintain periodic boundary conditions (in 
some volume $V$) for the bosonic fields involved, so the 
topological charges $Q$ are integers. Moreover we consider models 
with parity invariance. This implies $\la Q \ra = 0$, and the 
topological susceptibility is given by
\be
\chi_{\rm t} = \frac{1}{V} \la Q^{2} \ra \ .
\ee
In this framework, we are going to test the BCNW approximation
\cite{BCNW}. It can be written in
the form of an expansion in inverse powers of $V \chi_{\rm t}$,
\be
\la {\cal O} \ra_{Q} \simeq \la {\cal O} \ra + \frac{1}{V \chi_{\rm t}} 
c + \frac{1}{(V \chi_{\rm t})^{2}} (\bar c - c \, Q^{2} ) - 
\frac{2}{(V \chi_{\rm t})^{3}} \bar c \, Q^{2} \ .  
\label{approx3}
\ee
The left-hand-side refers to the expectation value of some
observable ${\cal O}$ (Refs.\ \cite{BCNW} inserted specifically 
the pion mass) within the sectors of topological charges $\pm Q$. 
It is accessible even in simulations which are confined 
to one --- or a few --- topological sectors. 

All the unknown terms on the right-hand-side, {\it i.e.}\ 
the expectation value $\la {\cal O} \ra$, $\chi_{\rm t}$ and 
the coefficients $c$ and $\bar c$, are quantities that asymptotically
stabilize in 
large volume. Hence this form enables the use of results for 
$\la {\cal O} \ra_{Q}$, measured in 
several volumes and for distinct $|Q|$, to determine these unknown terms. 
In particular we are interested in $\la {\cal O} \ra$ 
and $\chi_{\rm t}$. The coefficients are determined as well;
for instance $c$ can be expressed by derivatives with respect to
the vacuum angle $\theta$ of the extended Euclidean action
$S + \ri \theta Q$,
\be  \label{thetaeq}
c = \frac{1}{2} \la {\cal O} \ra ''(\theta )|_{\theta = 0} \ ,
\ee
but we are not going to discuss any conceivable interpretation
of these coefficients.

Actually the third order in approximation (\ref{approx3}) is incomplete,
but the additional term in this order would bring along another 
free parameter. These terms are identified and discussed in 
detail in Refs.\ \cite{Arthur13,Arthur14,CDWPol}.
Here we mostly focus on the simplest form which captures 
the $Q$-dependence of $\la {\cal O} \ra_{Q}$, and which
involves only three parameters (though an incomplete
second order),
\be  \label{approx2}
\la {\cal O} \ra_{Q} \approx \la {\cal O} \ra
+ \frac{c}{V \chi_{\rm t}} \left( 1 - \frac{Q^{2}}
{V \chi_{\rm t}} \right) \ .
\ee
In the following, we will refer to this approximation as the
{\em BCNW formula.}
Obviously we cannot determine the quantities $\la {\cal O} \ra$, 
$\chi_{\rm t}$ and $c$ within a single volume; for instance
\be  \label{fixedV}
\langle {\cal O} \rangle_{Q_{1}} - \langle {\cal O} \rangle_{Q_{2}} 
\approx \frac{c}{(V \chi_{\rm t})^{2}} (Q_{2}^{2} - Q_{1}^{2})
\ee
only determines the ratio $c / \chi_{\rm t}^{2}$. If we include different 
volumes $V_{1}$ and $V_{2}$, however, we could use {\it e.g.}\ 
$\langle {\cal O} \rangle_{0}(V_{1}) - \langle {\cal O} \rangle_{0}(V_{2})
\approx \frac{c}{\chi_{\rm t}} (1/V_{1} - 1/V_{2})$ to fix
$c/ \chi_{\rm t}$, and we obtain --- along with relation
(\ref{fixedV}) --- all three quantities, $\langle {\cal O} \rangle$,
$\chi_{\rm t}$ and $c$ (we repeat that only the former two are of interest).
In practice one would rather involve several volumes and topological 
sectors, and perform a 3-parameter fit to the (over-determined) system.

We distinguish three regimes for the volume $V$
\begin{itemize}

\item {\em Small volume:} there are significant finite size effects
of the ordinary type, not related to topology fixing, 
in particular in $\langle {\cal O} \rangle$ and $\chi_{\rm t}$.

\item {\em Moderate volume:} ordinary finite size effects are negligible
(they tend to be exponentially suppressed), but 
$\langle {\cal O} \rangle_{Q}$ still depends significantly on $|Q|$ and $V$.

\item {\em Large volume:} there are hardly any finite size effects left,
even the correction terms in approximations (\ref{approx3}), 
(\ref{approx2}) are negligible.

\end{itemize}

In small volumes, the formulae (\ref{approx3}) and
(\ref{approx2}) cannot be applied, 
because results from various volumes cannot be used for the same 
fit.\footnote{An extension of the BCNW approximation (\ref{approx2}) 
{\em including} ordinary finite size effects has been derived in Refs.\ 
\cite{PolProc}. This extension can be used for fits to data obtained 
from small volumes. It involves, however, additional fitting parameters.}
In large volumes, we obtain the correct value for 
$\langle {\cal O} \rangle$ anyhow, without worrying about
frozen topology, as we see from the expansions (\ref{approx3}) and
(\ref{approx2}). However, such large volumes may be inaccessible in 
realistic simulations, due to limitations of the computational resources.
Hence we are interested in {\em moderate volumes,} where the determination
of $\langle {\cal O} \rangle$ is difficult, but possibly feasible
by means of the BCNW approximation. Moreover, that regime also 
provides an estimate for $\chi_{\rm t}$, which is particularly
hard to measure directly.\\ 

The derivation of formula (\ref{approx3}) involves approximations,
which assume:\footnote{For convenience, this formula has been re-derived 
in Subsection 5.2 of Ref.\ \cite{BHSV} in a way, which highlights 
the r\^{o}le of these two assumptions.}

\begin{itemize}

\item $\langle Q^{2} \rangle = V \chi_{\rm t}$ is {\em large.}
As we mentioned before, eq.\ (\ref{approx3}) takes the form of
an expansion in $1/\langle Q^{2} \rangle$. Once $\chi_{\rm t}$ is 
stable, this can also be viewed as a large volume expansion. 

\item $| Q | / \langle Q^{2} \rangle $ is {\em small,} so we 
should work in the sectors with a small absolute value
$|Q|$. This is less obvious from the formulae (\ref{approx3})
and (\ref{approx2}) (although the terms $\propto Q^{2}$ are 
related to this condition), but it is required for a step in its 
derivation, which relies on a stationary phase approximation.

\end{itemize}

Here we employ numerical data to explore
how large $\langle Q^{2} \rangle$ has to be for this 
approximation to be sensible, and up to which absolute
value $|Q|$ the data are useful in this context.
In practice it is rather easy to work at small $|Q|$, but 
the former condition could be a serious obstacle.

So far there have been only few attempts to apply this approximation 
to simulation data. This was done for the 2-flavor 
Schwinger model with dynamical overlap fermions \cite{BHSV,WBIH}
with respect to the pseudo-scalar mass $M_{\pi}$ and the 
chiral condensate $\Sigma$.
Tests for a quantum rotor ---  more precisely a scalar particle 
on a circle with a potential --- are reported in 
Refs.\ \cite{Arthur13,Arthur14}.

Another approach was derived --- similarly to the BCNW 
approximation --- in Ref.\ \cite{AFHO}. It refers to the 
long-distance correlation of the topological charge density
$q(x)$, $Q = \int d^{d}x \, q(x)$. 
The applicability of that method has been tested in a set of
models \cite{topdense}, and variants had been studied
previously \cite{AFHOvari}.
Further approaches to extract physics from topologically frozen
Markov chains include Refs.\ \cite{LSD,Liao,slab}.
Preliminary results of this work have been anticipated in some
proceeding contributions \cite{Arthur13,CDWPol,PolProc,MexFrankprocs}.

\section{Tests for the quantum rotor}

As a simple but precise test, we first consider a toy model
from quantum mechanics ({\it i.e.}\ 1d quantum field theory), namely 
the quantum rotor, or 1d XY model, or 1d O(2) model. 
It describes a free quantum mechanical particle moving on
a circle, with a periodicity condition in Euclidean time.
A theoretical discussion of this system, in the continuum 
and for different lattice actions, is given in 
Ref.\ \cite{rot97}.\footnote{For the analytic treatment,
Ref.\ \cite{rot97} uses the Hamiltonian formalism. A discussion 
in terms of path integrals is given in Ref.\ \cite{Schulman}.} 
Below we write down the continuum action, 
and on the lattice the standard action and 
the Manton action \cite{Manton} (in lattice units),
\bea
S_{\rm cont} [\vp ] & = & \frac{\beta_{\rm cont}}{2} \int_{0}^{L_{\rm cont}} 
dt \, \dot \vp (t)^{2} \ , \nn \\
S_{\rm standard} [\vp ] & = & \beta \sum_{t=1}^{L}
\Big( 1 - \cos ( \Delta \vp_{t} ) \Big) \ , \nn \\
S_{\rm Manton}[\vp ] & = & \frac{\beta}{2} \sum_{t=1}^{L}
\, ( \Delta \vp_{t} )^{2} \ .
\label{rotactions}
\eea
$L_{\rm cont}$ and $L$ are the extent of the periodic Euclidean time 
interval in the continuum and on the lattice, respectively,
$\vp (t)$ and $\vp_{t}$ are time dependent angles, with
$\vp(L_{\rm cont}+t) = \vp (t)$, $\vp_{L+t} = \vp_{t}$.
$\beta_{\rm cont}$ and $\beta$ can be interpreted as an 
inverse temperature,
or in this case also as the moment of inertia.
In the terms for the lattice actions we define
\be
\Delta \vp_{t} = (\vp_{t+1} - \vp_{t}) \ {\rm mod} \ 2 \pi 
\in ( - \pi, \pi ] \ ,
\ee
{\it i.e.}\ the modulo function is implemented such that
it minimizes $| \Delta \vp_{t}|$. Thus $\Delta \vp_{t}$ also
defines the lattice topological charge density $q_{t}$ (geometric 
definition) and the charge $Q$,
\be
q_{t} = \frac{1}{2 \pi} \Delta \vp_{t} \ , \quad
Q [\vp ] = \sum_{t=1}^{L} q_{t} \in \Z \ .
\ee

In the continuum and infinite size $L_{\rm cont}$, the correlation length 
and its product with the topological susceptibility amount to
\be  \label{contxichi}
\xi_{\rm cont} = 2 \beta_{\rm cont} \ , \quad 
\chi_{\rm t} \ \xi_{\rm cont} = \frac{1}{2 \pi^{2}} \ .
\ee
Analytic expressions for the corresponding quantities on the 
lattice, with the standard action and the Manton action, are 
given in Ref.\ \cite{rot97}.

Our simulations were carried out with the Wolff cluster algorithm 
\cite{Ulli}, which performs non-local update steps. This algorithm
is highly efficient and provided a statistics of $5 \times 10^{9}$
measurements for each setting. Since it changes the topological 
sector frequently, in this case the observables could also be 
measured directly to high precision, which allows for a detailed 
test of the BCNW method.
In most quantum field theoretic models no efficient cluster
algorithm is known, in particular in the presence of gauge fields.
Then one has to resort to local update algorithms, which motivates
this project, as we pointed out in Section 1.

For our tests we set $\beta = 4$ and consider six lattice sizes in the
range $L= 150 \dots 400$. 
This is large compared to the correlation length, which was measured 
at $L=400$ as
\be
\xi_{\rm standard} = 6.81495(4) \ , \quad
\xi_{\rm Manton} = 7.9989(1) \ ,
\ee
very close to the analytic values at $L = \infty$.
This demonstrates that ordinary finite size effects are very small,
but --- as we are going to see --- there are significant fixed topology 
finite size effects. Hence we are in the regime of moderate volumes,
as desired. Moreover, this regime is sensible also because
lattice artifacts are quite well suppressed.

The BCNW formula consists of leading terms in an expansion in
$1/\la Q^{2} \ra$, cf.\ Section 1. In the range $L=150 \dots 400$
we obtain
\be  \label{Q21dO2}
\la Q^{2} \ra_{\rm standard} = 1.13 \dots 3.02 \ , \quad
\la Q^{2} \ra_{\rm Manton} = 0.95 \dots 2.53 \ .
\ee
This suggests that we are in the transition regime to the validity
of this method, which is interesting to explore.

\subsection{Action density}

We first consider the action density 
\be  \label{actdens}
s = \la S \ra /V \ . 
\ee
This quantity is not directly physical, but it is
suitable for testing the BCNW method, based on topologically 
restricted expectation values $s_{|Q|} = \la S \ra_{|Q|}/V$. 
Moreover, the corresponding fits provide a value for $\chi_{\rm t}$,
which {\em is} physical.

\begin{figure}[h!]
\begin{center}
\hspace*{-6mm}
\includegraphics[width=0.375\textwidth,angle=270]{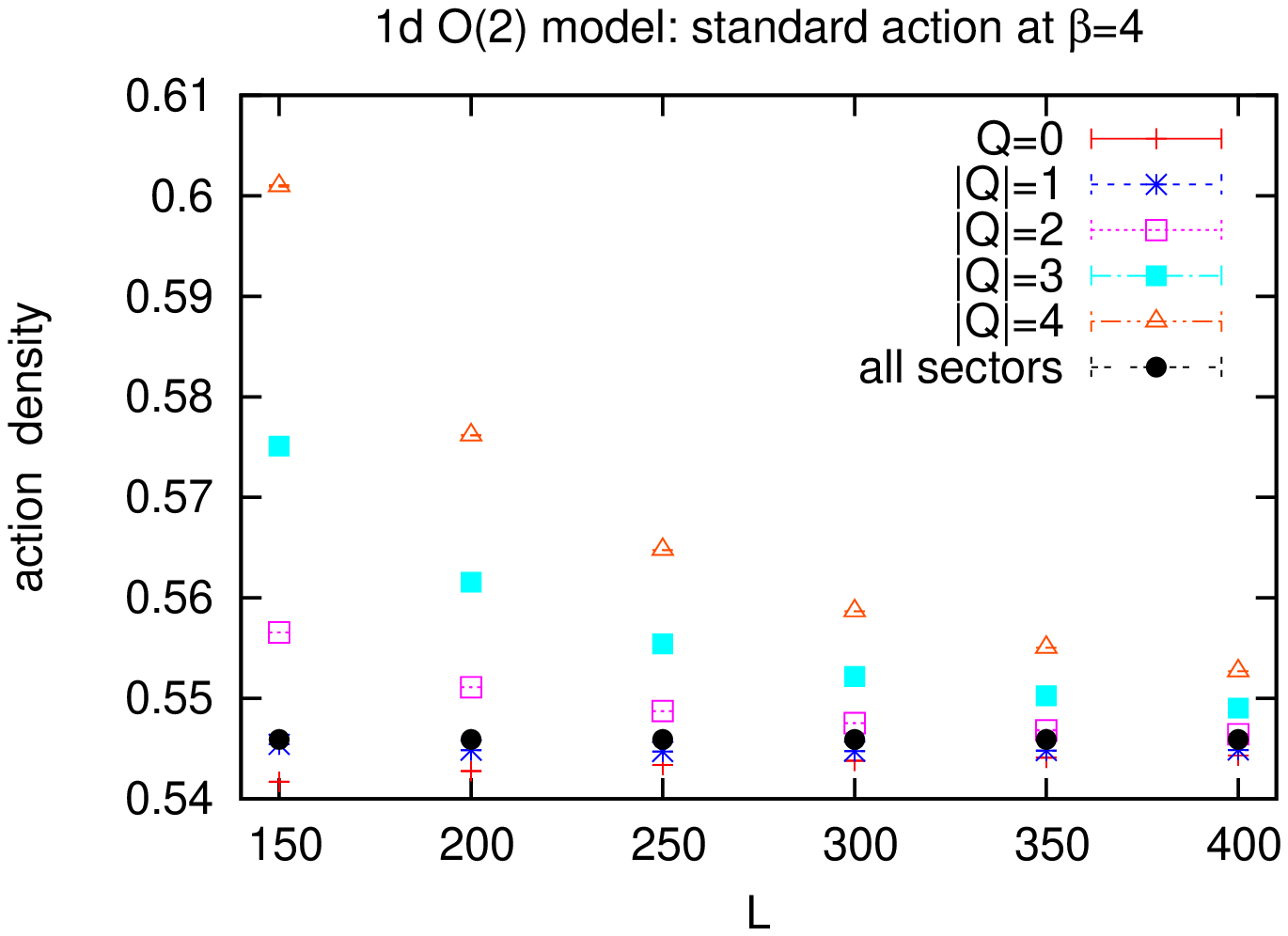}
\hspace*{-8mm}
\includegraphics[width=0.375\textwidth,angle=270]{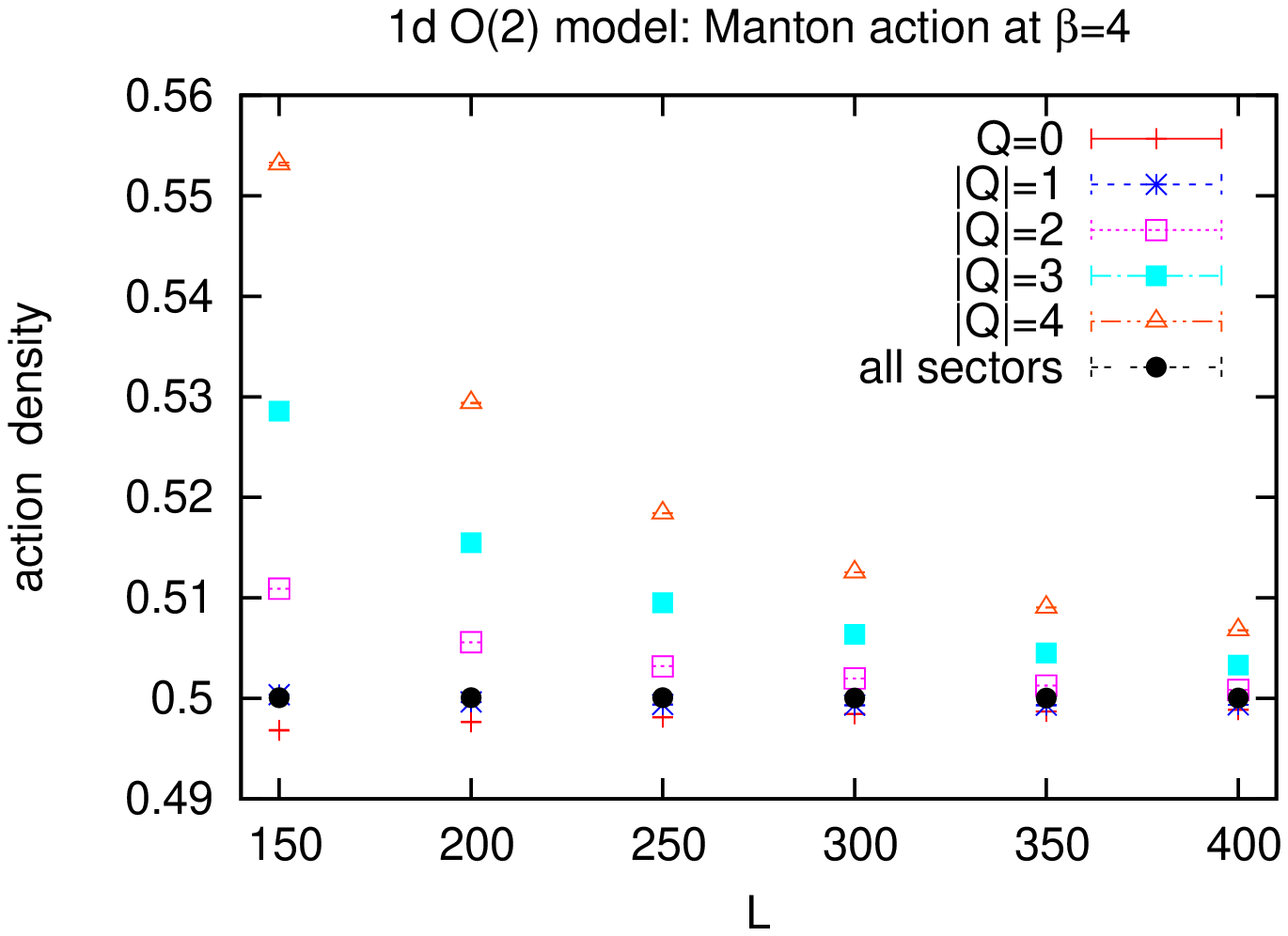} 
\vspace*{-4mm}
\caption{The action density in the 1d O(2) model at $\beta =4$
on lattices of size $L=150 \dots 400$, with the standard action (left) 
and the Manton action (right). We show $s$ measured in all sectors 
(which is practically constant in this range of $L$), as well as 
the values of $s_{|Q|}$ in the sectors $|Q|= 0 \dots 4$, which
strongly depend on $|Q|$ and $V$.}
\vspace*{-3mm}
\label{actden1dO2}
\end{center}
\end{figure}

Figure \ref{actden1dO2} shows the action density for both lattice
actions under consideration, measured at fixed $|Q| = 0 \dots 4$,
and by including all sectors (the way the simulation samples them). 
The latter is constant to high accuracy for $L=150
\dots 400$, which confirms that ordinary finite size effects are
negligible. On the other hand, at fixed $|Q|$ we see
deviations far beyond the statistical errors, depending on $L$
and $|Q|$, so this setting is appropriate for the application
of the BCNW method.

Table \ref{actden1dO2tab} presents our results obtained by least-square
fits to the BCNW approximation (\ref{approx2}): we use data for
$s_{|Q|}$ in all six volumes, and in the topological sectors
$|Q|=0 \dots |Q|_{\rm max}$, where $|Q|_{\rm max}$ varies from 1 to 4.
Similar results are obtained when we only involve the larger
volumes, such as $L=250 \dots 400$ or $300 \dots 400$.
\begin{table}[ht!]
\begin{center}
\begin{tabular}{|c||c|c||c|c|}
\hline
& \multicolumn{2}{|c||}{standard action} & \multicolumn{2}{|c|}{Manton action} \\
\hline
$|Q|_{\rm max}$ & $s$ & $\chi_{\rm t}$ & $s$ & $\chi_{\rm t}$ \\
\hline
 1  & 0.545910(1) & 0.007552(4) &  0.500073(3) & 0.006135(9) \\ 
 2  & 0.545910(1) & 0.007555(3) &  0.500072(2) & 0.006132(8) \\
 3  & 0.545912(2) & 0.007559(5) &  0.500072(2) & 0.006132(8) \\
 4  & 0.545912(2) & 0.007559(5) &  0.500072(2) & 0.006131(7) \\
\hline
all & 0.545910(1) & 0.007554\hspace*{5mm}  & 0.500041(1) & 0.006333~~~~ \\
\hline
\end{tabular}
\caption{Results based on fits to the formula (\ref{approx2}), 
with input data for the action density
in the range $L= 150 \dots 400$ and $|Q| \leq |Q|_{\rm max}$.
The last line displays $s$ measured in all sectors at $L=400$,
and the analytic value of $\chi_{\rm t}$ at $L=\infty$.}
\label{actden1dO2tab}
\end{center}
\vspace*{-5mm}
\end{table}

Regarding the value of $s$, the method works perfectly (to the given
precision) for the standard action, and up to a deviation of about
$0.006 \, \%$ for the Manton action.
For the standard action the fits yield values for $\chi_{\rm t}$, which  
are again compatible with the correct value, with uncertainties around
$0.05 \, \%$. In case of the Manton action a systematic discrepancy 
of $3 \, \%$ is observed, as a consequence of the approximations in
formula (\ref{approx2}). 

In summary, this first numerical experiment can be considered a 
success of the BCNW method. The good results for $s$ are highly 
non-trivial in view of the sizable differences in the individual sectors 
(shown in Figure \ref{actden1dO2}), and exactly these differences
give rise to quite good estimates for $\chi_{\rm t}$.
As a generic property, it is easy to measure $s_{|Q|}$ accurately
(in gauge theories it is given by the mean plaquette value),
so it is motivated to estimate $\chi_{\rm t}$ in this way
also in higher dimensional models.

\subsection{Magnetic susceptibility}

In this model, the correlation function in a fixed sector
of topological charge $Q$ has a peculiar form. For a continuous
time variable $t$ it reads \cite{Arthur14}
\bea  \label{corrfun}
\la \vec e (0) \cdot \vec e (t) \ra_{Q} &=& \frac{1}{2} \exp \Big( - 
\frac{t (L_{\rm cont}-t)}{2 \beta_{\rm cont} L_{\rm cont}} \Big) 
\cos \Big( \frac{2 \pi Qt}{L_{\rm cont}} \Big) \ , \\
{\rm with} && \vec e (t) = \left( \begin{array}{c}
\cos \vp (t) \\ \sin \vp (t) \end{array} \right) \ . \nn
\eea
The unusual last factor in eq.\ (\ref{corrfun}) obstructs the 
determination of a correlation length $\xi_{Q \neq 0}$,
and we recall that the BCNW method does not apply to results,
which are obtained in various volumes, but always at $Q=0$.

By integrating over the time shift $t$, however, we obtain 
a quantity, which is suitable for testing this method, namely
the magnetic susceptibility 
\be  \label{magsus}
\chi_{\rm m} = \frac{\la \vec M^{2} \ra - \la \vec M \ra^{2}}{L_{\rm cont}} 
= \int_{0}^{L_{\rm cont}} dt \, \la \vec e (0) \cdot \vec e (t) \ra -
\frac{1}{L_{\rm cont}} 
\Big( \Big\la \int_{0}^{L_{\rm cont}} dt \, \vec e (t) \Big\ra \Big)^{2} \ ,
\ee
where $\vec M = \int_{0}^{L_{\rm cont}} dt \, \vec e (t)$ is the magnetization.
The subtracted term vanishes in our case due to the global O(2) 
invariance, $\la \vec M \ra = \vec 0$.
The magnetic susceptibility is physical in the framework of 
statistical mechanics; we can interpret a configuration $[\vec e \, ]$ 
as a spin chain. Based on eq.\ (\ref{corrfun}) we obtain for its 
topologically restricted counterpart
\be  \label{intchim}
\chi_{{\rm m},|Q|}
= 2 \int_{0}^{L_{\rm cont}/2} dt \, \exp \Big( - \frac{t}{2 \beta_{\rm cont}} 
+ \frac{t^{2}}{2 \beta_{\rm cont} L_{\rm cont}} \Big) \, 
\cos \Big (\frac{2 \pi Q t}{L_{\rm cont}} \Big) \ . 
\ee
In each sector, the limit $L_{\rm cont} \to \infty$ leads to 
$\chi_{\rm m} = \chi_{{\rm m},|Q|} = 4 \beta_{\rm cont}$.
If we insert the large volume expansions
of $\exp (t^{2}/(2 \beta_{\rm cont} L_{\rm cont}))$ and 
$\cos (2 \pi Q t /L_{\rm cont})$ up to ${\cal O}(1/L_{\rm cont}^{3})$,
and perform the integral, we arrive at
\bea
\chi_{{\rm m},Q} & = & \chi_{\rm m} +
\frac{4 \beta_{\rm cont}}{\pi^{2} L_{\rm cont} \chi_{\rm t}} \Big( 1 
+ \frac{3/ \pi^{2} - Q^{2}}{L_{\rm cont} \chi_{\rm t}} \Big) \nn \\
&& + \frac{12 \beta_{\rm cont}}{\pi^{4} (L_{\rm cont} \chi_{\rm t})^{3}} 
\Big( \frac{5}{\pi^{2}} - 2 Q^{2} \Big) + {\cal O} \Big( \frac{1}
{(L_{\rm cont} \chi_{\rm t})^{4}} \Big) \ ,
\label{BCNWchim}
\eea
where we substituted the infinite volume value 
$\chi_{\rm t} = 1/(4 \pi^{2} \beta_{\rm cont})$ \cite{rot97}, cf.\ 
eq.\ (\ref{contxichi}).\footnote{The finite size effects in $\chi_{\rm t}$, 
and those due to the upper bound of the integral in eq.\ (\ref{intchim}), 
are exponentially suppressed.}
This is exactly the form of the BCNW approximation (\ref{approx3}),
with 
\be  \label{ccbar}
c = \frac{4 \beta_{\rm cont}}{\pi^{2}} \ , \quad
\bar c = \frac{12 \beta_{\rm cont}}{\pi^{4}} \ ,
\ee
and in this case the third order is complete.
If we only consider the second order and neglect its $\bar c$-term, 
we are left with the BCNW approximation (\ref{approx2}).
A detailed derivation of the expansion (\ref{BCNWchim}) is given
in Appendix A.

Therefore the magnetic susceptibility is fully appropriate for
numerical tests of the validity of this approximation, where we 
use the  corresponding lattice terms, like 
$\vec M = \sum_{t=1}^{L} \vec e_{t}$. The sources of
systematic errors (errors in the BCNW approximation) are
sub-leading finite size effects and lattice artifacts.

\begin{figure}[h!]
\begin{center}
\hspace*{-6mm}
\includegraphics[width=0.375\textwidth,angle=270]{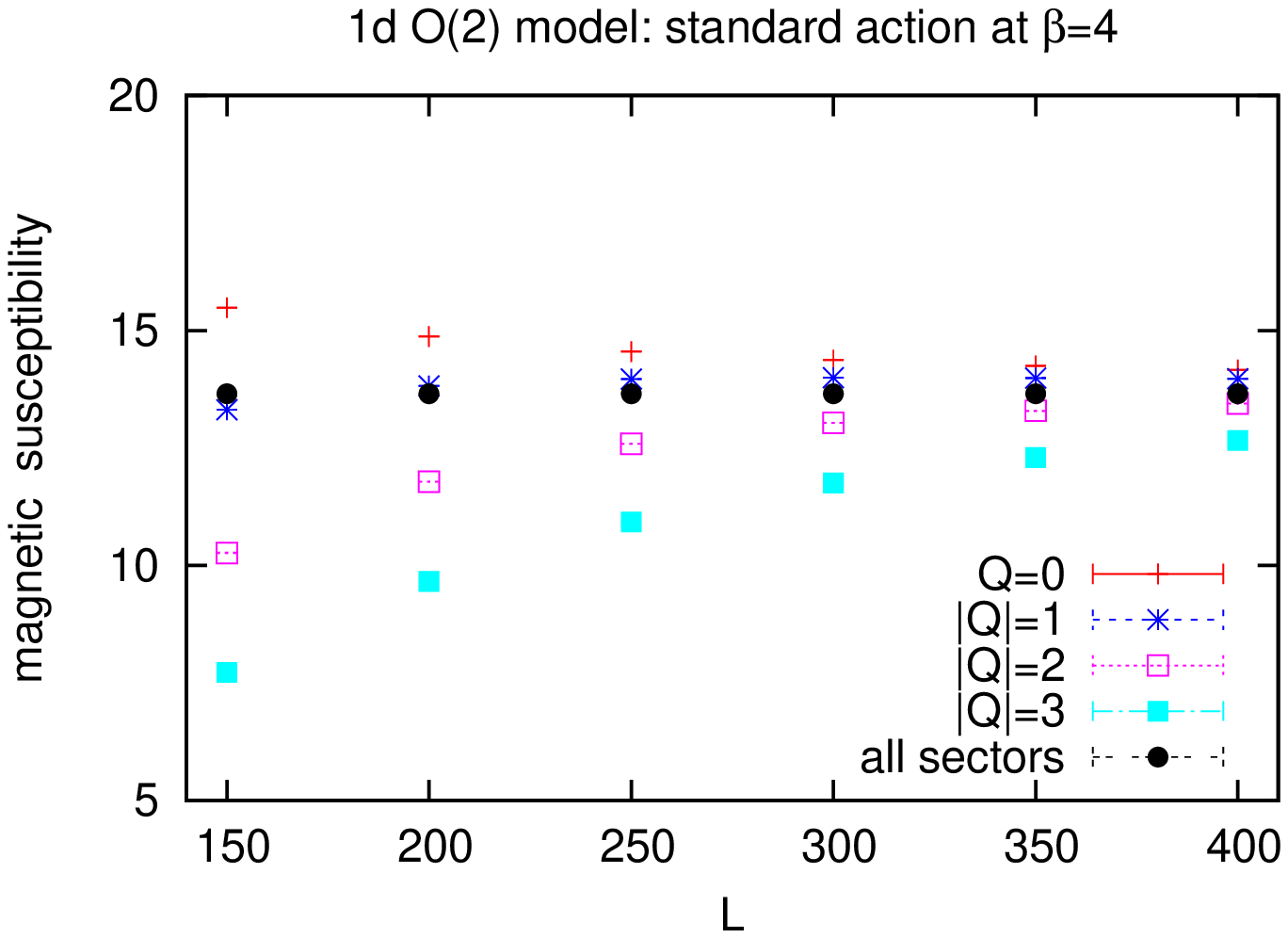}
\hspace*{-8mm}
\includegraphics[width=0.375\textwidth,angle=270]{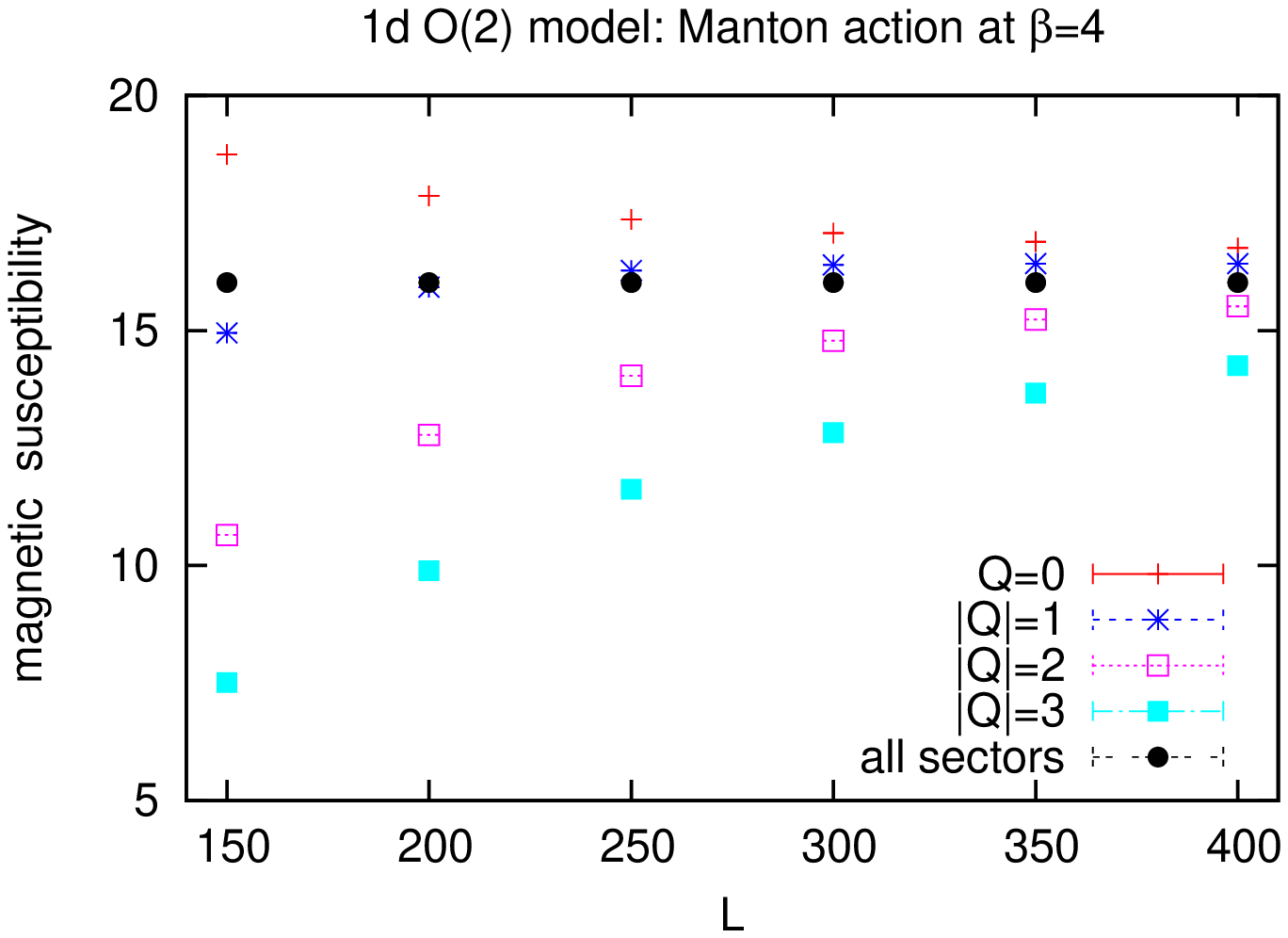} 
\vspace*{-4mm}
\caption{The magnetic susceptibility in the 1d O(2) model at $\beta =4$
on lattices of size $L=150 \dots 400$, with the standard action (left) 
and the Manton action (right). We show $\chi_{\rm m}$ measured in all
sectors (practically constant in this range of $L$), 
as well as $\chi_{{\rm m},|Q|}$ in the sectors $|Q|= 0 \dots 3$
(well distinct).}
\vspace*{-3mm}
\label{magsus1dO2}
\end{center}
\end{figure}

In analogy to Subsection 3.1, Figure \ref{magsus1dO2} gives an
overview over the values of $\chi_{{\rm m},|Q|}$ up to $|Q|=3$,
at different $L$. Again we see that the value measured in all
sectors is stable in $L$, whereas the topologically restricted
results strongly depend on $L$ and $|Q|$. Hence the setting
is suitable for the BCNW method also with respect to the magnetic
susceptibility.

We proceed to the fits to search the optimal values --- according
to formula (\ref{approx2}) --- for the (over-determined) 
susceptibilities $\chi_{\rm m}$ and $\chi_{\rm t}$. Table \ref{magsus1dO2tab}
shows the results in the fitting ranges $L = L_{\rm min} \dots 400$,
$L_{\rm min} = 150,$ $250$, $300$,
and $|Q| = 0 \dots |Q|_{\rm max}$, with $|Q|_{\rm max} =2$ or $3$.

\begin{table}[ht!]
\begin{center}
\begin{tabular}{|c|c||l|l||l|l|}
\hline
& & \multicolumn{2}{|c||}{standard action} & 
\multicolumn{2}{c|}{Manton action} \\
\hline
$L_{\rm min}$ & $|Q|_{\rm max}$ & 
\multicolumn{1}{|c|}{$\chi_{\rm m}$} & 
\multicolumn{1}{|c||}{$\chi_{\rm t}$} & 
\multicolumn{1}{|c|}{$\chi_{\rm m}$} & 
\multicolumn{1}{|c|}{$\chi_{\rm t}$} \\
\hline
150 & 2 & 13.64(16)  & 0.0072(13) &  16.11(35)  & 0.0054(18) \\ 
150 & 3 & 13.67(22)  & 0.0070(22) &  16.14(41)  & 0.0050(26) \\ 

250 & 2 & 13.64(5)   & 0.0071(5)  &  16.00(14)  & 0.0060(8) \\ 
250 & 3 & 13.65(13)  & 0.0074(15) &  15.99(28)  & 0.0064(20) \\ 

300 & 2 & 13.64(5)   & 0.0071(5)   &  16.02(12)  & 0.0058(8) \\ 
300 & 3 & 13.66(13)  & 0.0073(17)  &  16.02(29)  & 0.0061(23) \\ 
\hline
 & all  & 13.6545(4) & 0.007554   &  16.0187(5) & 0.006333 \\
\hline
\end{tabular}
\caption{Results based on fits to formula (\ref{approx2}), with input
data for the magnetic susceptibility
in the range $L= L_{\rm min} \dots 400$ and $|Q| \leq |Q|_{\rm max}$.
The last line displays $\chi_{\rm m}$ measured in all sectors at $L=400$,
and $\chi_{\rm t}$ at $L=\infty$.}
\label{magsus1dO2tab}
\end{center}
\vspace*{-5mm}
\end{table}

The fitting results for both susceptibilities are compatible 
with the correct values, albeit the uncertainty of $\chi_{\rm t}$
is rather large. Without knowing the exact value one could combine
the results of separate fits, which reduces the uncertainty, but it
leads to a $\chi_{\rm t}$-value which is somewhat too small.
On the other hand, for $\chi_{\rm m}$ the values are far more precise,
and the relative uncertainty is on the percent level (or below) in
each case. Here a combination which reduces the uncertainty is 
welcome, although it has to be done with care since
the partial results are not independent of each other.
We add that the fitting results for the coefficient $c$ are 
consistent with eq.\ (\ref{ccbar}), $c \simeq 1.6$, within 
(considerable) uncertainties.

The observed precisions for $\chi_{\rm m}$ and $\chi_{\rm t}$
can be understood if we consider the impact of the sub-leading 
contributions, which are missing in the BCNW formula (\ref{approx2}): 
taking into account the additional terms up to the incomplete 
third order modifies
the fitting results for $\chi_{\rm m}$ only on the permille level, but
those for $\chi_{\rm t}$ in ${\cal O}(10) \, \%$, both with somewhat
enhanced errors. Also a variety of further fitting variants,
with the terms of a complete second or complete third order of
approximation (\ref{BCNWchim}), with fixed or free additional terms, 
leads to consistent results for $\chi_{\rm m}$ and $\chi_{\rm t}$, but 
with enlarged errors. In summary, there seems to be no fitting
strategy which improves the results compared to the simple 3-parameter
fit based on the BCNW approximation (\ref{approx2}).

\section{Applications to the 2d Heisenberg model}

Our study of the 2d Heisenberg model, or 2d O(3) model, uses quadratic 
lattices of unit spacing and square-shaped volumes $V = L \times L$. 
On each lattice site $x$ there is a classical spin $\vec e_{x} \in S^{2}$,
and we implement periodic boundary conditions in both directions. 
We consider the standard lattice action as well as the constraint 
action \cite{topact},
\bea
S [\vec e \, ]_{\rm standard} &=& \beta \sum_{x,\mu} (1 - \vec e_{x} \cdot
\vec e_{x + \hat \mu}) \ , \nn \\
S [\vec e \, ]_{\rm constraint} &=& \left\{ \begin{array}{ccc}
0 & & \vec e_{x} \cdot \vec e_{x + \hat \mu} \geq \cos \delta \quad 
\forall x, \, \mu = 1,2 \\
+ \infty && {\rm otherwise,} \end{array} \right.  
\label{act2dO3}
\eea
where $\delta$ is the constraint angle, and $\hat \mu$ is the
unit vector in $\mu$-direction.

Our simulations were performed 
at $\beta =1.5$ and $\delta = 0.55 \, \pi$, respectively,
with the correlation lengths
\bea
{\rm standard~action} \  
\ (L=84) &:&
\xi = 9.42(2) \ , \nn \\ 
{\rm constraint~action}
\ (L = 96) &:&
\xi = 3.58(5) \ .
\eea
The cluster algorithm allowed us to perform ${\cal O}(10^{7})$
measurements at each lattice size shown
in Figures \ref{actden2dO3} and \ref{2dO3constraint}.

For the topological charge we use again a geometric definition
\cite{BergLuscher}. To this end, each plaquette is split into
two triangles, in alternating orientation. We consider the
oriented solid angle of the spins at the corners of a triangle:
the sum of the two angles (divided by $4\pi$) within a plaquette
(associated with the site $x$) amounts to its
topological charge density $q_{x}$.
Due to the periodic boundary conditions,
their sum must be an integer, $Q = \sum_{x} q_{x} \in \Z$.
Details and explicit formulae are given in Refs.\ \cite{topact,topdense}.

\subsection{Action density}

\begin{figure}[h!]
\begin{center}
\vspace*{-3mm}
\includegraphics[width=0.53\textwidth,angle=270]{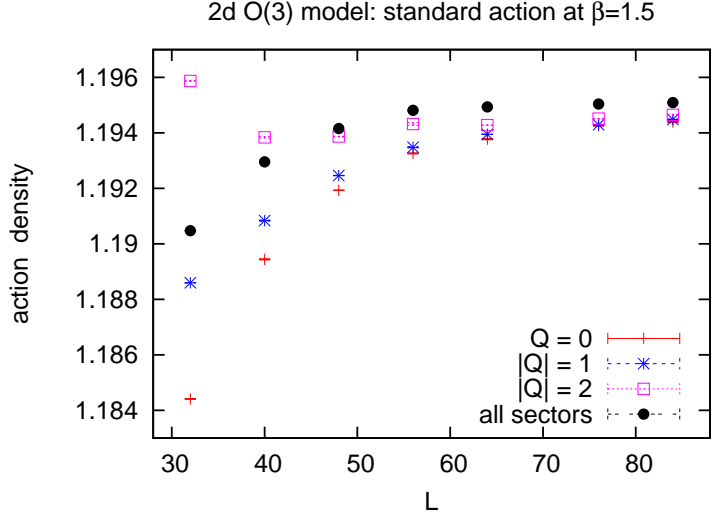}
\vspace*{-1mm}
\caption{The action density in the 2d O(3) model, on $L \times L$ lattices 
with the standard lattice action, in the sectors with topological
charge $|Q| = 0$, $1$, $2$, and summed over all sectors 
({\it i.e.}\ all configurations used for the numerical measurements).
The latter stabilizes to $0.3$ permille for $L \geq 56$.}
\vspace*{-3mm}
\label{actden2dO3}
\end{center}
\end{figure}

A study of the BCNW formula with respect to the action density 
(\ref{actdens}) can only be performed with the standard action (in 
case of the constraint action all contributing configurations have 
action $S_{\rm constraint}=0$). 
Figure \ref{actden2dO3} shows the values of $s$ and $s_{|Q|}$,
$|Q| \leq 2$ in the range $L=32 \dots 84$. The total expectation
value $s$ is stable within $0.0003$ 
for $L \geq 56$, while the topologically constrained results differ 
by ${\cal O}(10^{-3})$ even at $L=84$. Therefore $L = 56 \dots 84$ is a 
regime of moderate volumes, which is suitable for testing the BCNW formula.

The fitting results, for $|Q| \leq 2$ and various ranges of $L$ 
are listed in Table \ref{actden2dO3tab}. The fits do not match
the BCNW formula perfectly, as expected, since the 
latter is an approximation, and the input data have very small
statistical errors of ${\cal O} (10^{-5})$.\footnote{Of course,
the ratio $\chi^{2}/$d.o.f.\ could be reduced by adding more
terms to the $1/V$-expansion. However, in Table \ref{2dO3tab} we
are going to demonstrate that
this does not improve the results for the observable and for 
$\chi_{\rm t}$, in qualitative agreement with Section 3.}
Nevertheless, the value of $s$ is obtained correctly up to a high
precision of $0.2$ permille. On the other hand, the
determination of the topological susceptibility is less successful;
only the fit with $L = 76$ and $84$ yields a result, which is
correct within the errors.

\begin{table}[ht!]
\begin{center}
\begin{tabular}{|c||l|l|c|}
\hline
fitting range in $L$ & \multicolumn{1}{|c|}{$s$} & 
\multicolumn{1}{|c|}{$\chi_{\rm t}$} & $\chi^{2}/$d.o.f. \\
\hline
56 --- 64 & 1.1955(2)  & 0.0035(5) & 2.66 \\
56 --- 76 & 1.19538(6) & 0.0031(3) & 2.66 \\
56 --- 84 & 1.19536(5) & 0.0030(3) & 2.63 \\
\hline
64 --- 76 & 1.19532(7) & 0.0031(3) & 2.65 \\
64 --- 84 & 1.19531(5) & 0.0031(3) & 2.58 \\
76 --- 84 & 1.1953(1)  & 0.0026(3) & 2.60 \\
\hline
\hline
$L=84$, all sectors & 1.195089(5) & 0.002323(3) & \\ 
\hline
\end{tabular}
\caption{Fitting results for the action density $s$ and the topological
susceptibility $\chi_{\rm t}$ in the 2d O(3) model. The input data
in fixed topological sectors are plotted in Figure \ref{actden2dO3}.}
\label{actden2dO3tab}
\end{center}
\vspace*{-5mm}
\end{table}

\subsection{Magnetic susceptibility and correlation length}

\begin{figure}[h!]
\begin{center}
\includegraphics[width=0.53\textwidth,angle=270]{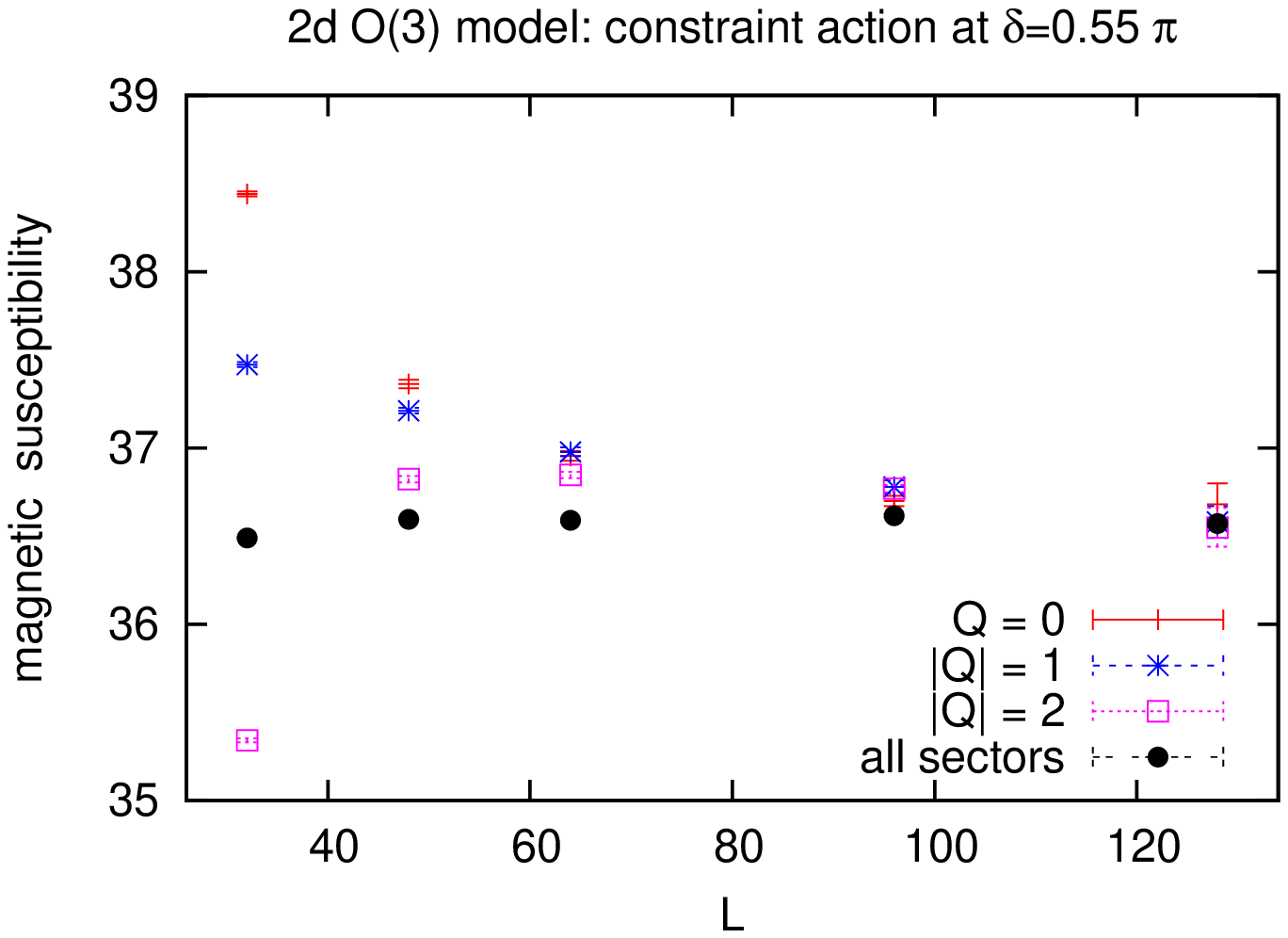}
\includegraphics[width=0.53\textwidth,angle=270]{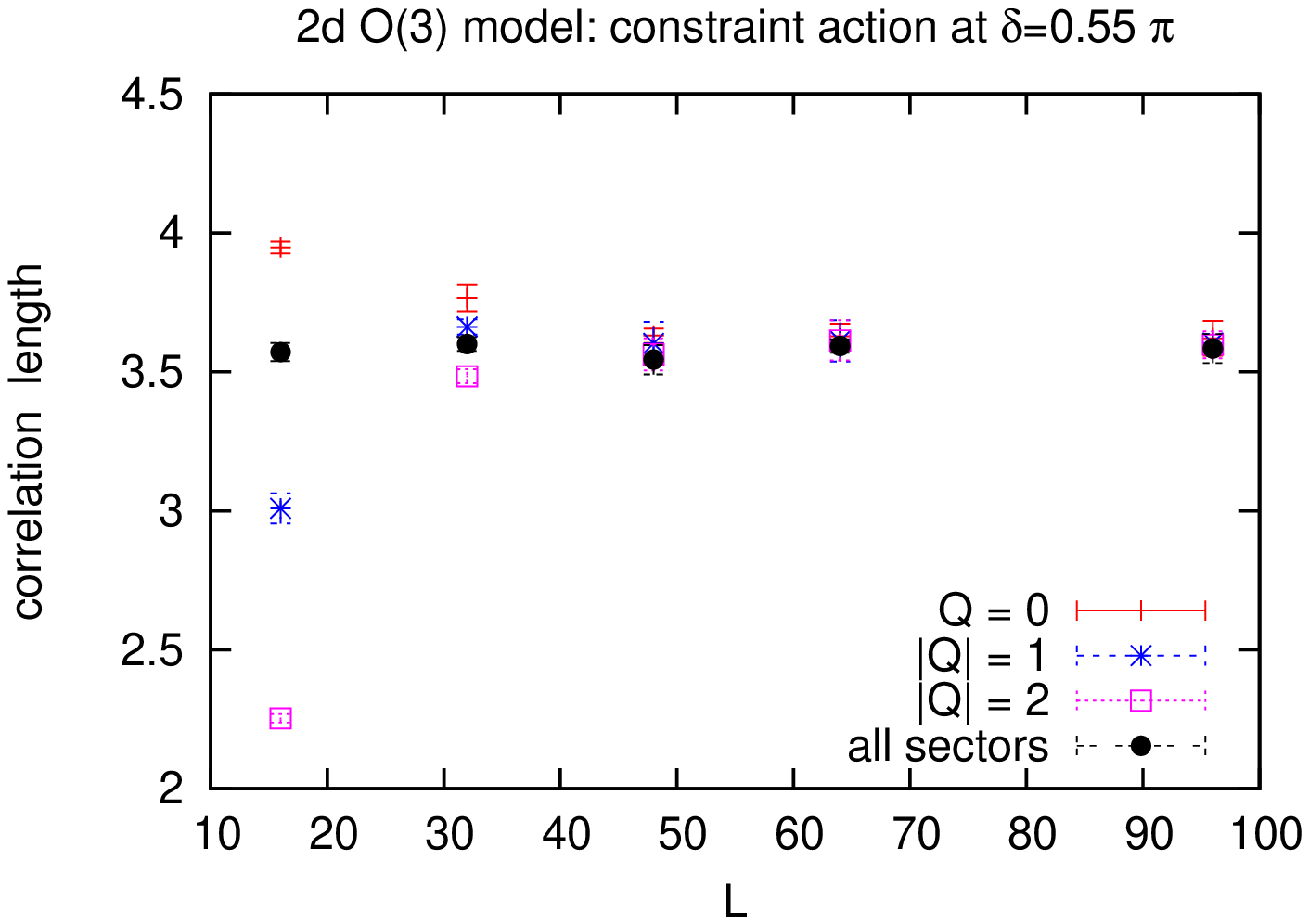} 
\vspace*{-2.5mm}
\caption{Results for the magnetic susceptibility (above) and for the
correlation length (below) in the 2d O(3) model, with the constraint
action at $\delta = 0.55 \, \pi$. The windows, which are 
suitable for applications of the BCNW formula,
are given by $L=48 \dots 96$ for $\chi_{\rm m}$, and by $L=32 \dots 64$ 
for $\xi$.}
\vspace*{-5mm}
\label{2dO3constraint}
\end{center}
\end{figure}

We proceed to the constraint action (\ref{act2dO3}) 
where our choice of $\delta$ yields a shorter correlation length,
which favors the stabilization of observables (measured in all sectors)
at smaller $L$. This can be seen in Figure \ref{2dO3constraint},
which shows the magnetic susceptibility $\chi_{\rm m}$, analogous 
to eq.\ (\ref{magsus})
(again the disconnected part vanishes due to rotational symmetry),
and the correlation length $\xi$. Stabilization within the errors
is attained for $\chi_{\rm m}$ at $L\geq 48$
(with errors around $0.2$ permille), 
and for $\xi$ already at $L \geq 16$ 
(with errors of ${\cal O}(1) \, \%$). 
On the other hand, for $L = 128$ the $\chi_{{\rm m},|Q|}$-values
are not distinguished anymore from $\chi_{\rm m}$ beyond the errors,
and the same happens for $\xi_{|Q|}$ already at $L=96$. Finally, we
have to exclude $L=16$, because here we only obtain $\la Q^{2} \ra
\simeq 0.63$, hence its inverse is not suitable as an expansion
parameter. This singles out the regime of moderate volumes, 
where the BCNW formula is appropriate,
to the range $L= 48 \dots 96$ for $\chi_{\rm m}$, 
and $L= 32 \dots 64$ for $\xi$. 

Our fitting results are given in Table \ref{2dO3tab}.
In the case of $\chi_{\rm m}$ we probe the
BCNW formula (\ref{approx2}) (with its incomplete second order,
${\cal O}(1/V^{2})$), as well as its extensions to the second order 
plus an incomplete third order as given in formula (\ref{approx3}). 
For the latter option, the approximation is more precise, 
but an additional free parameter $\bar c$ hampers the fits.

\begin{table}[ht!]
\begin{center}
\begin{tabular}{|c|c||c|c||c|}
\hline
 & fitting & BCNW    & incomplete  & all sectors \\
 & range   & formula & 3rd order   & at $L_{\rm max}$ \\
\hline
\hline
$\chi_{\rm m}$ & \multirow{2}{*}{48 --- 64}
& 36.56(4)   & 36.64(11) & 36.590(9) \\
$\chi_{\rm t}$ & & 0.0026(2) & 0.0031(6) & 0.0027935(14) \\
\hline
$\chi_{\rm m}$ & \multirow{2}{*}{48 --- 96}
& 36.58(3) & 36.64(7) & 36.616(9) \\
$\chi_{\rm t}$ & & 0.0026(2) & 0.0032(6) & 0.0027942(11) \\
\hline
$\xi$ & \multirow{2}{*}{32 --- 64} & 3.56(2) & 3.58(4) & 3.59(2) \\
$\chi_{\rm t}$ & & 0.0027(3) & 0.0034(14) & 0.0027935(14) \\
\hline
\end{tabular}
\caption{Fitting results based on data for $\chi_{\rm m}$ and for $\xi$ 
in the 2d O(3) model, in fitting ranges $L_{\rm min}$ --- $L_{\rm max}$,
and sectors with $|Q| \leq 2$. 
In the case of $\chi_{\rm m}$, with the optimal range, 
we show results for the BCNW approximation
(\ref{approx2}), as well as its extension to the complete second order 
plus one term of ${\cal O}(1/V^{3})$, according to formula (\ref{approx3}).}
\label{2dO3tab}
\end{center}
\vspace*{-3mm}
\end{table}

For both fitting versions, the results for $\chi_{\rm m}$
and $\chi_{\rm t}$ are compatible with the directly measured values.
We observe, however, that the inclusion of terms 
beyond the BCNW formula enhances 
the uncertainty (due to the additional fitting parameter).
The uncertainty is on the permille level for $\chi_{\rm m}$, but
large for $\chi_{\rm t}$, in particular with extra terms.
(Without these terms it is around $ 8 \, \%$.)
It turns out to be non-profitable to extend the
approximation beyond the BCNW formula.

The simple BCNW approximation is also superior for the fits 
with respect to $\xi$, where the additional terms drastically 
increase the uncertainty.
The results in Table \ref{2dO3tab} are correct, within 
percent level for $\xi$, but again with a large 
uncertainty of the $\chi_{\rm t}$-value.

We add that we also tried fits to the complete second order
approximation, without the third order term that appears in
formula (\ref{approx3}). However, this scenario (which also
involves the fitting parameter $\bar c$) is clearly unfavorable:
in this case, it often happens that the least-square fit even fails 
to converge to values in the correct magnitude.

To conclude, this study suggests that the simple BCNW formula, 
with only three free parameters, is in fact optimal for extracting 
values for the considered observable, and for $\chi_{\rm t}$.
Moreover, we confirm that the method works best for the 
determination of the observable; it is less successful with 
respect to the determination of $\chi_{\rm t}$.

\section{Results in 4d SU(2) Yang-Mills theory}

The topological tunnelling rate has been investigated in 4d 
Yang-Mills theories with the heatbath \cite{4dYMheatbatch} and 
the HMC algorithm \cite{4dYMHMC}. In both cases the autocorrelation 
time with respect to 
$Q$ was found to 
increase drastically for decreasing lattice spacing, which further 
substantiates the motivation of our study.

\subsection{\label{SEC001}Simulation setup}

We consider 4d SU(2) Yang-Mills theory, which has the continuum action
\begin{equation}
S_{\rm cont} [A] = \beta_{\rm cont} \int d^{4}x \, 
F_{\mu \nu}^{a} (x) F_{\mu \nu}^{a} (x) \ ,
\end{equation}
and the topological charge
\begin{equation}
Q[A] = \frac{1}{16 \pi^{2}} \int d^{4}x \, 
\epsilon_{\mu \nu \rho \sigma} F_{\mu \nu}^a (x) F_{\rho \sigma}^a (x) \ .
\end{equation}

On the lattice we simulate Wilson's standard plaquette action.
For the topological charge of a lattice gauge configuration $[U]$,
we use an improved field-theoretic definition \cite{Forcrand97},
\begin{equation}
\label{EQN001} Q[U] = \frac{1}{16 \pi^2} \sum_x
\epsilon_{\mu \nu \rho \sigma} \sum_{\Box=1,2,3} \frac{c_\Box}{\Box^4} 
F_{x,\mu \nu}^{(\Box \times \Box)}[U] F_{x,\rho \sigma}^{(\Box \times \Box)}[U]  \ ,
\end{equation}
where $F_{x,\mu \nu}^{(\Box \times \Box)}[U]$ denotes the
lattice field strength tensor, clover averaged over square-shaped loops 
of size $\Box \times \Box$, and $(c_{1},c_{2},c_{3}) = (1.5, -0.6, 0.1)$.
Before applying eq.\ (\ref{EQN001}), we perform a number of cooling 
sweeps with the intention to remove local fluctuations in the gauge 
configurations, while preserving the topological structure. 

\begin{figure}[htb!]
\vspace*{-2mm}
\begin{center}
\includegraphics[width=0.67\textwidth]{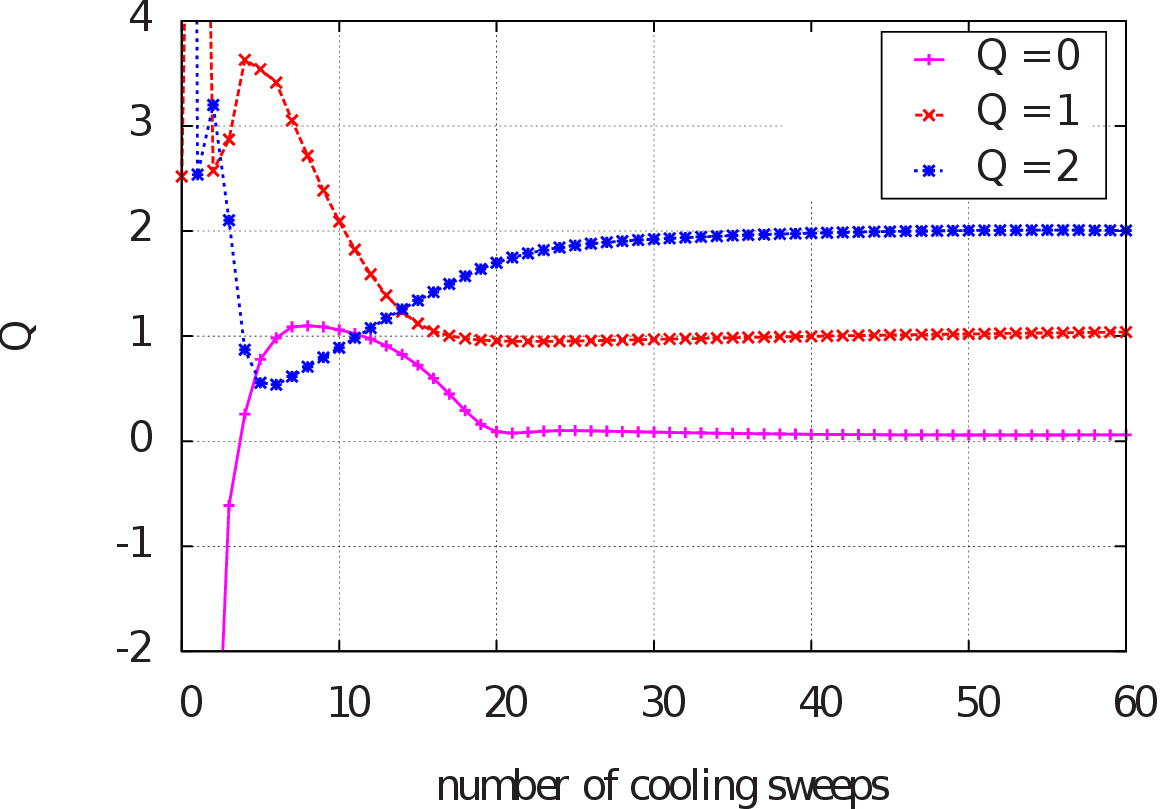}
\vspace*{-2mm}
\caption{\label{FIG051}Cooling and assignment of the topological charge 
for three typical gauge configurations, at $\beta = 2.5$, 
in a lattice volume $V = 18^{4}$.}
\end{center}
\vspace*{-5mm}
\end{figure}
A cooling sweep amounts to a local minimization of the action, 
{\it i.e.}\ a minimization with respect to each gauge link 
within a short range. For this minimization we use again an 
improved lattice Yang-Mills action,
\begin{equation}
\label{EQN002} S[U] = \frac{\beta}{16} \sum_x \sum_{\mu \nu} 
\sum_{\Box=1,2,3} \frac{c_\Box}{\Box^4} \textrm{Tr} 
\Big(\uno - W_{x,\mu \nu}^{(\Box \times \Box)}[U]\Big) \ ,
\end{equation}
where $W_{x,\mu \nu}^{(\Box \times \Box)}[U]$ is a clover averaged loop of 
size $\Box \times \Box$ with the coefficients $c_\Box$ given
above (for comparison, the standard plaquette action corresponds
to $(c_{1},c_{2},c_{3}) = (1,0,0)$). Choosing an appropriate number 
of cooling sweeps is a subtle and somewhat ambiguous task, which 
is carried out for each gauge configuration one by one. 
After every cooling sweep we compute $Q[U]$ according 
to eq.\ (\ref{EQN001}). As soon as $Q[U]$ is stable (it varies by less 
than 10\,\% and is close to an integer for at least 50 cooling sweeps), 
the corresponding close integer is the topological charge that we assign 
to the gauge configuration $[U]$. Figure~\ref{FIG051} shows examples 
for typical cooling histories of gauge configurations with $Q = 0$, 
$1$ and $2$. (Details of this procedure, and a comparison to other 
definitions of the topological charge, are discussed in 
Ref.\ \cite{CichyLat14}.)

Our simulations were performed with a heatbath algorithm,
see {\it e.g.}\ Ref.\ \cite{Creutz}. 
We set $\beta = 2.5$, which corresponds to a lattice 
spacing $a \approx 0.073 \, \textrm{fm}$, if the scale is
set with the QCD Sommer parameter $r_0 = 0.46 \, \textrm{fm}$ 
\cite{PhilipsenWagner}. This value
is in the range of lattice spacings 
$0.05 \, \textrm{fm} \ltapprox a \ltapprox 0.15 \, \textrm{fm}$ 
typically used in contemporary QCD simulations. We generated 
gauge configurations in lattice volumes $V=L^{4}$, with 
$L = 12, \, 14 , \, 15 , \, 16 , \, 18$.\footnote{Unless stated 
otherwise, we continue using lattice units.} In each volume, 
observables were measured on 4000 configurations, separated by 
100 heatbath sweeps. This guarantees their statistical independence;
in particular, even the auto-correlation time with respect to the
topological charge $Q$ is below 20 heatbath sweeps.

\subsection{\label{SEC010}Computation of observables}

The observable we focus on is the static quark-antiquark 
potential ${\mathcal V}_{q \bar{q}}(r)$ for 
separations $r = 1, 2 \ldots 6$. This quantity 
can be interpreted as the mass of a static-static meson. 
To determine ${\mathcal V}_{q \bar{q}}(r)$, we consider temporal 
correlation functions of operators
\begin{equation}
O_{q \bar{q}} (r) = \bar{q}(\vec r_{1}) \, U^\textrm{APE}
(\vec r_{1}, \vec r_{2}) \, q(\vec r_{2}) \ , \quad 
r = |\vec r_{1} - \vec r_{2}| \ ,
\end{equation}
where $\bar{q}$, $q$ represent spinless static quarks, while
$U^\textrm{APE}(\vec r_1 , \vec r_2 )$ denotes a product of APE 
smeared spatial links \cite{APE} along a straight line connecting 
the lattice sites $\vec r_1$ and $\vec r_2$ on a given time slice. 
For the quarks we use the HYP2 static action, which is designed 
to reduce UV fluctuations and, therefore, to 
improve the signal-to-noise ratio \cite{HYP2}. 
These temporal correlation functions can be simplified analytically 
resulting in Wilson loop averages $\langle W(r,t) \rangle$ 
with APE smeared spatial and HYP2 smeared temporal lines of 
length $r$ and $t$, respectively. Thus we arrive at the vacuum 
expectation value
\begin{equation}
\la \Omega | O^\dagger_{\bar{q} q}(t) O_{\bar{q} q}(0) | \Omega \ra 
\propto \Big\langle W(r,t) \Big\rangle \ .
\end{equation}
We chose the APE smearing parameters as $N_\textrm{APE} = 15$ and 
$\alpha_\textrm{APE} = 0.5$, which (roughly) optimizes the overlap of 
$O_{\bar{q} q} | \Omega \rangle$ with the ground state of the static 
potential (for details of the smearing procedure we refer to
Ref.\ \cite{ETM}, where a similar setup had been used).

\subsection{Numerical results}

\subsubsection{\label{SEC387}The static potential}

\begin{figure}[htb]
\vspace*{-2mm}
\begin{center}
\includegraphics[width=0.7\textwidth]{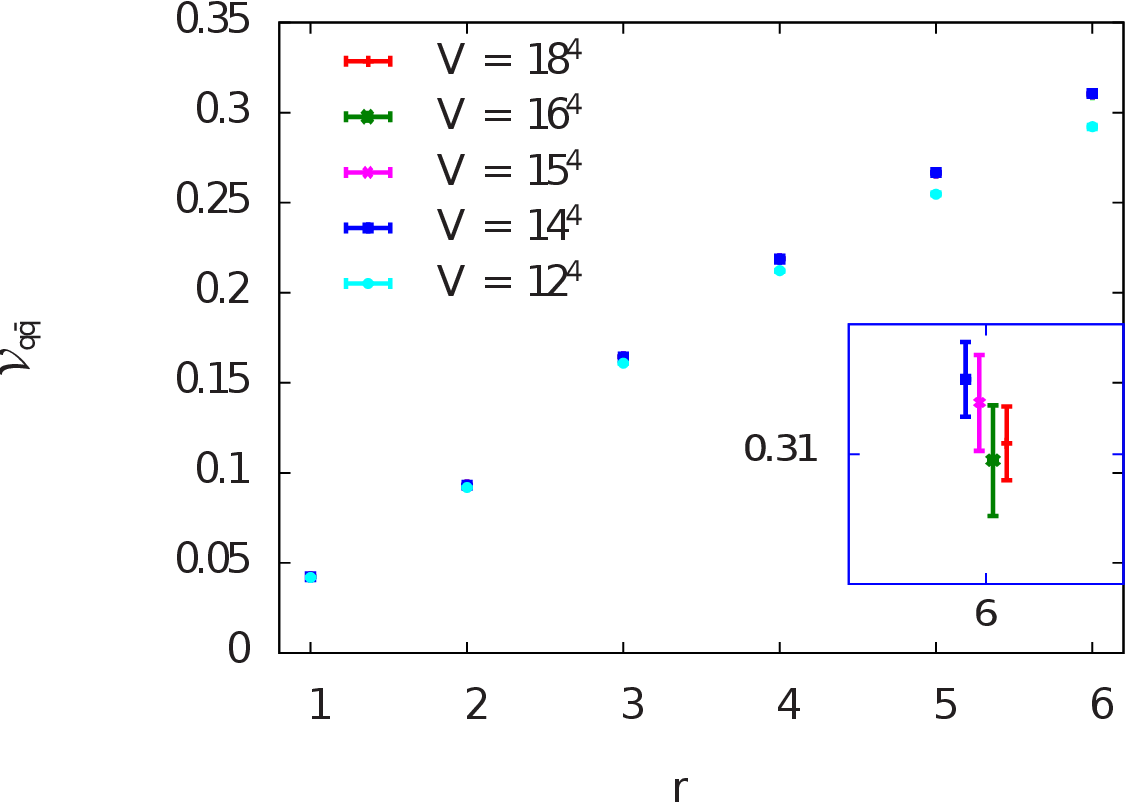}
\vspace*{-2mm}
\caption{\label{FIG001}The static potential ${\mathcal V}_{q \bar{q}}(r)$ 
in a variety of lattice volumes $V = 12^{4} \dots 18^4$.}
\end{center}
\vspace*{-5mm}
\end{figure}

Figure~\ref{FIG001} shows results for the static potential 
measured in all topological sectors, {\it i.e.}\ for each $r$ 
and $t$ the Wilson loop average is computed on all configurations, 
which are available in some volume.\footnote{As usual,
we determined ${\mathcal V}_{q \bar{q}}(r)$ by searching for
a plateau value of the effective mass
$m_\textrm{eff}(r,t) = \log(\la W(r,t+1) \ra / \la W(r,t) \ra)$.} 
The volumes 
$14^4$, $15^4$, $16^4$ and $18^4$ yield identical results within 
statistical errors, but the static potential in the 
$12^4$ volume differs by several $\sigma$ for quark-antiquark 
separations $r \geq 3$. We conclude that $V = 12^4$ 
entails sizable ordinary finite volume effects (not 
associated with topology fixing), whereas for volumes 
$V \geq 14^4$ such ordinary finite volume effects are 
negligible. Consequently, we do not 
use the $12^4$ lattice in the following fixed topology 
studies.\footnote{We repeat that the BCNW formula can be extended
by incorporating ordinary finite volume effects \cite{PolProc}.}

\begin{figure}[htb]
\vspace*{-2mm}
\begin{center}
\includegraphics[width=0.67\textwidth]{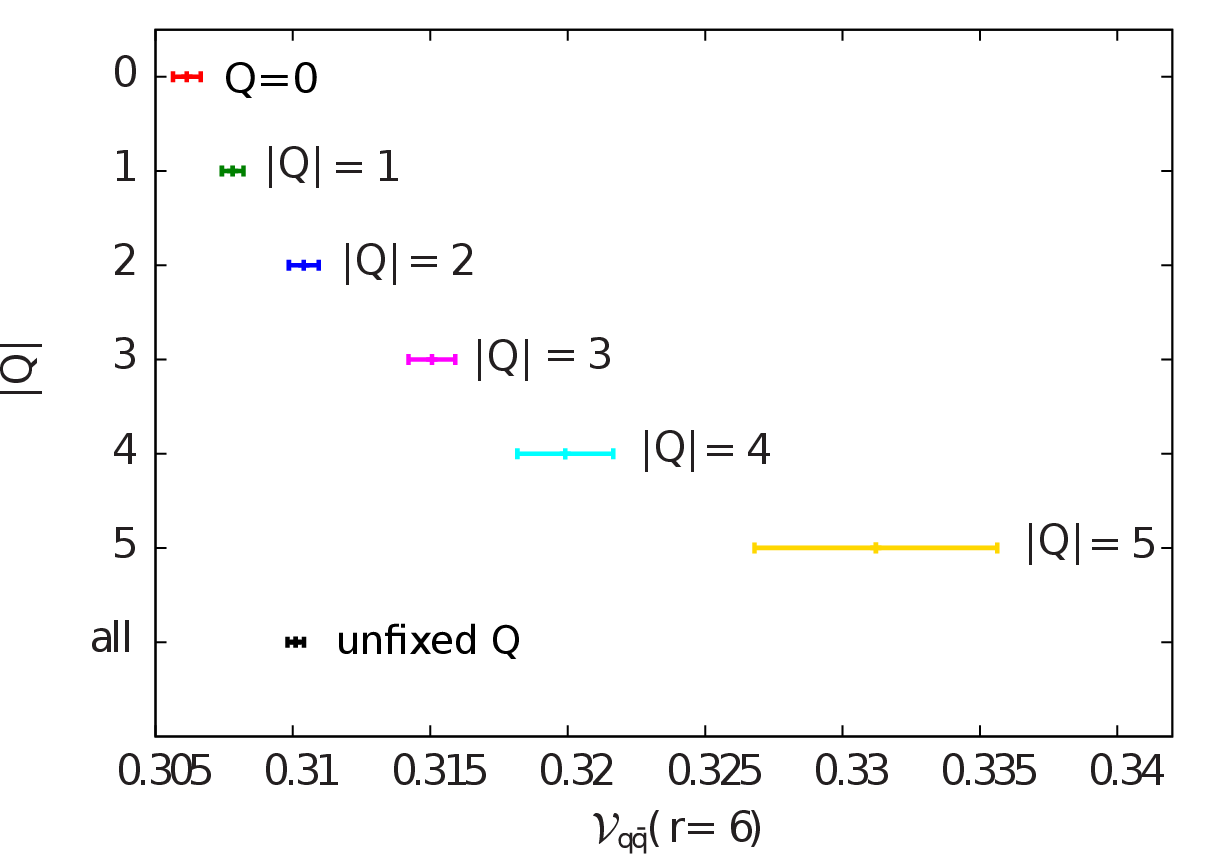}
\vspace*{-2mm}
\caption{\label{FIG002}The static potential at separation
$r=6$, ${\mathcal V}_{q \bar{q}}(6)$, for fixed topological 
sectors $|Q| \leq 5$, and without topology 
fixing, in the volume $V = 15^{4}$.}
\end{center}
\vspace*{-5mm}
\end{figure}

For $V = 15^4$, Figure~\ref{FIG002} demonstrates that 
static potentials obtained at fixed topology from different 
sectors $|Q| = 0 \ldots 5 $ 
(by averaging only over configurations of a fixed charge $|Q|$),
${\mathcal V}_{q \bar{q}, |Q|}$, differ
significantly.\footnote{Again we determined 
${\mathcal V}_{q \bar{q}}(r)$ by fitting constants to effective 
mass plateaux. Even though topology has been fixed, the effective 
masses exhibit a constant behavior (within statistical errors)
at large $t$.} For example 
${\mathcal V}_{q \bar{q},0}(r=6)$ and ${\mathcal V}_{q \bar{q},4}(r=6)$
differ by more than $6 \sigma$. They are also well distinct 
from the corresponding result in all sectors, 
${\mathcal V}_{q \bar{q},|Q|\leq 1}(6) < {\mathcal V}_{q \bar{q}}(6)
< {\mathcal V}_{q \bar{q}, |Q| \geq 2}(6)$. 
These observations show that $V = 14^{4} \dots 18^{4}$ is in the
regime that we denoted as {\em moderate volumes} (cf.\ Section 2),
where the BCNW method is appropriate to extract
observables from fixed topology measurements.
Similar results for the static potential in SU(3) Yang-Mills theory 
have been reported in Ref.\ \cite{Bruck10}.

To extract the physical static potential from Wilson loop 
averages, separately computed in distinct topological sectors 
$|Q| \leq 7$ and some volume $V$, $\la W_{V}(r,t) \ra_{|Q|}$, 
we follow the procedure discussed in Ref.\ \cite{Arthur13}.
\begin{itemize}
\item We perform $\chi^2$ minimizing fits of either the $1/V$ 
expansion of the correlation function \cite{BCNW},
\begin{eqnarray}
\nonumber & & \hspace{-0.7cm} C_{Q,V}(t) = 
\Big\la W_{V}(r,t) \Big\ra_{|Q|} \\
\label{EQN003} & & \approx \alpha(r) 
\exp\bigg\{ -\bigg[ {\mathcal V}_{q \bar{q}}(r) + \frac{1}{2} 
{\mathcal V}''_{q \bar{q}}(r) \frac{1}{V \chi_{\rm t}} 
\bigg(1 - \frac{Q^2}{V \chi_{\rm t}}\bigg)\bigg] t\bigg\}
\end{eqnarray}
(cf.\ formula (\ref{approx2})), or of the improved 
approximation \cite{Arthur14}
\begin{eqnarray}
\nonumber & & \hspace{-0.7cm} C_{Q,V}(t) \simeq 
\frac{\alpha(r)}{\sqrt{1 + 
{\mathcal V}''_{q \bar{q}}(r) t / (\chi_{\rm t} V)}} \\
\label{EQN004} & & \hspace{0.675cm} \times \exp\bigg(
-{\mathcal V}_{q \bar{q}}(r) t - \frac{1}{\chi_{\rm t} V} 
\bigg(\frac{1}{1 + {\mathcal V}''_{q \bar{q}}(r) t / 
(\chi_{\rm t} V)} - 1\bigg) \frac{1}{2} Q^2\bigg) \qquad \quad 
\end{eqnarray}
with respect to the parameters ${\mathcal V}_{q \bar{q}}(r)$, 
${\mathcal V}''_{q \bar{q}}(r) 
= \partial^{2}_{\theta}{\mathcal V}_{q \bar{q}}(r, \theta)|_{\theta = 0} $, 
$\alpha(r)$ ($r = 1 \ldots 6$) 
and $\chi_{\rm t}$ to the numerical results for 
$\la W_{V}(r,t) \ra_{|Q|}$ in the range 
$t_\textrm{min} \leq t \leq t_\textrm{max}$, where $t_\textrm{min}$ 
and $t_\textrm{max}$ are displayed in Table~\ref{TAB580}.\footnote{Again,
$\theta$ is the vacuum angle that we referred to before 
in eq.\ (\ref{thetaeq}).} 
When fitting formula (\ref{EQN004}), we 
also study the scenario where $\chi_{\rm t}$ is fixed to 
$\chi_{\rm t} = 7 \times 10^{-5}$, which was obtained in 
Ref.\ \cite{Forcrand97} by means of a direct measurement, in 
agreement with the fixed topology study in Ref.\ \cite{topdense}.
Moreover, we checked that the resulting fit 
parameters are stable within errors when we vary $t_\textrm{min}$ 
and $t_\textrm{max}$ by $\pm 1$.

\begin{table}[htb]
\centering
\begin{tabular}{|c|c|c|c|c|}
\hline 
$V$ & $t_\textrm{min}$ & $t_\textrm{max}$ & maximum $|Q|$ 
fulfilling & maximum $|Q|$ fulfilling \\
 & & & $1 / (\chi_{\rm t} V) , \, |Q|/(\chi_{\rm t} V) < 1$ 
& $1 / (\chi_{\rm t} V) , \, |Q|/(\chi_{\rm t} V) < 0.5$ \\
\hline 
$14^4$ & 5 & 7 & 2 & 1 \\
$15^4$ & 5 & 7 & 3 & 1 \\
$16^4$ & 5 & 8 & 4 & 2 \\
$18^4$ & 5 & 8 & 7 & 3 \\
\hline 
\end{tabular}
\caption{\label{TAB580}Temporal fitting ranges $t_\textrm{min}
\ldots t_\textrm{max}$, and maximum topological charges $|Q|$, for 
the lattice volumes $V$ under consideration.}
\end{table}

\item The results for $\la W_{V}(r,t) \ra_{|Q|}$ 
entering the fit are restricted to those $|Q|$ and $V$ values for 
which $1 / (\chi_{\rm t} V) , \, |Q|/(\chi_{\rm t} V) < 1$ or $< 0.5$;
we recall that the approximations (\ref{EQN003}) and (\ref{EQN004}) 
are only valid for sufficiently large 
$\chi_{\rm t} V = \la Q^{2} \ra$, and small $|Q|$.
To implement this selection we insert $\chi_{\rm t} = 7 \times 10^{-5}$
\cite{Forcrand97}; Table~\ref{TAB580} gives an overview.

\item We either perform a single combined fit to all considered 
separations $r = 1 \ldots 6$, or six separate fits, one 
for each $r$. In the latter case we obtain 
six results for $\chi_{\rm t}$, which agree within the errors in 
most cases, cf.\ Subsection~\ref{SEC588}.
\end{itemize}

Table~\ref{TAB001} collects the
results for ${\mathcal V}_{q \bar{q}}(r)$ from fixed topology 
computations (using four volumes, $V = 14^4 , \, 
15^4 , \, 16^4 , \, 18^4$), and computed in all sectors
at $V = 18^4$. There is agreement 
between most of these results within about $2 \sigma$.
Only for $r=1$, and the relaxed constraint 
$1 / (\chi_{\rm t} V) , \, |Q|/(\chi_{\rm t} V) < 1$, there are a few 
cases with discrepancies beyond $3 \sigma$, in particular for the 
expansion (\ref{EQN003}) (the corresponding data in Table~\ref{TAB001} 
are displayed in italics). 

The extent of the errors of the fitting results is fairly 
independent of the choice of the expansion ((\ref{EQN003}), or
(\ref{EQN004}), or (\ref{EQN004}) with $\chi_{\rm t} = 7 \times 10^{-5}$ 
fixed). The errors increase, however, by factors up to 
$\approx 2$, when we implement the stringent constraint 
$1 / (\chi_{\rm t} V) , \, |Q|/( \chi_{\rm t} V) < 0.5$, which is 
expected, since less input data are involved, see Table \ref{TAB580}. 
All fits of the expansions (\ref{EQN003}) and (\ref{EQN004})
capture well the fixed topology results.
\begin{table}[p]
\hspace*{-1cm}
\begin{tabular}{|c|c|c|c|c|c|c|}
\hline
method & $\mathcal{V}_{q \bar{q}}(1)$ & $\mathcal{V}_{q \bar{q}}(2)$ 
& $\mathcal{V}_{q \bar{q}}(3)$ & $\mathcal{V}_{q \bar{q}}(4)$ 
& $\mathcal{V}_{q \bar{q}}(5)$ & $\mathcal{V}_{q \bar{q}}(6)$
\tabularnewline
\hline
\hline
 \multicolumn{7}{|c|}{all sectors,$\quad$ $V =18^4$}
\tabularnewline
\hline
 & 0.04229(1) & 0.09329(2) & 0.1646(1) & 0.2190(1) & 0.2664(2) 
& 0.3101(3)\tabularnewline
\hline
\hline
 \multicolumn{7}{|c|}{fixed topology,$\quad$ $V \in 
\{ 14^4 , \, 15^4 , \, 16^4 , \, 18^4 \}$,$\quad$ $1 
/ (\chi_{\rm t} V) , \, |Q|/(\chi_{\rm t} V) < 1$}\tabularnewline
\hline
(\ref{EQN003})c & {\it 0.04240(3)}
& 0.09343(8) 
& 0.1646(2) & 0.2189(3) & 0.2662(4) & 0.3097(5)\tabularnewline
(\ref{EQN003})s & {\it 0.04241(3)} & 0.09342(9) 
& 0.1646(2) & 0.2189(3) & 0.2662(4) & 0.3097(6)\tabularnewline
\hline 
(\ref{EQN004})c & 0.04230(3) & 0.09324(8) & 0.1644(2) 
& 0.2187(3) & 0.2661(4) & 0.3098(6)\tabularnewline
(\ref{EQN004})s & {\it 0.04240(3)} & 0.09338(9) 
& 0.1645(2) & 0.2188(3) & 0.2661(4) & 0.3098(6)\tabularnewline
\hline 
(\ref{EQN004})c$\chi_{\rm t}$ & 0.04225(3) & 0.09326(8) & 0.1643(2) 
& 0.2186(3) & 0.2660(4) & 0.3097(6)\tabularnewline
(\ref{EQN004})s$\chi_{\rm t}$ & 0.04225(3) & 0.09326(8) & 0.1643(2) 
& 0.2186(3) & 0.2660(4) & 0.3097(6)\tabularnewline
\hline 
\hline
 \multicolumn{7}{|c|}{fixed topology,$\quad$ 
$V \in \{ 14^4 , \, 15^4 , \, 16^4 , \, 18^4 \}$,$\quad$ 
$1 / (\chi_{\rm t} V) , \, |Q|/(\chi_{\rm t} V) < 0.5$}
\tabularnewline
\hline
(\ref{EQN003})c & 0.04227(4) & 0.09326(14) & 0.1645(3) 
& 0.2190(5) & 0.2665(7) & 0.3103(10)\tabularnewline
(\ref{EQN003})s & 0.04226(4) & 0.09322(13) & 0.1644(3) 
& 0.2189(5) & 0.2666(8) & 0.3105(11)\tabularnewline
\hline 
(\ref{EQN004})c & 0.04227(4) & 0.09326(14) & 0.1645(4) 
& 0.2190(5) & 0.2665(7) & 0.3104(10)\tabularnewline
(\ref{EQN004})s & 0.04226(4) & 0.09323(13) & 0.1645(3) 
& 0.2189(5) & 0.2665(8) & 0.3104(10)\tabularnewline
\hline 
(\ref{EQN004})c$\chi_{\rm t}$ & 0.04225(4) & 0.09317(12) 
& 0.1643(3) & 0.2186(4) & 0.2660(6) & 0.3096(8)\tabularnewline
(\ref{EQN004})s$\chi_{\rm t}$ & 0.04225(3) & 0.09317(12) 
& 0.1643(3) & 0.2186(4) & 0.2660(6) & 0.3096(8)\tabularnewline
\hline
\end{tabular}
\caption{\label{TAB001}Results for the static potential 
${\mathcal V}_{q \bar{q}}(r)$ for separations 
$r = 1 \ldots 6$ measured with and without topology fixing. 
In the column ``method'' the equation number 
of the expansion is listed, ``c'' denotes a single {\em combined} 
fit for all separations $r= 1 \ldots 6$, ``s'' denotes a 
{\em separate} fit for each separation, and $\chi_{\rm t}$ indicates
that the topological susceptibility is not a fit parameter, but 
fixed to $\chi_{\rm t} = 7 \times 10^{-5}$.
Fixed topology results, which differ by more than $3 \sigma$ 
from the directly computed value, are written in italics.}
\end{table}

For the extraction of the potential it seems essentially
irrelevant whether a single combined fit or six separate 
fits are performed. Both the mean values and the statistical 
errors of ${\mathcal V}_{q \bar{q}}(r)$ are in most cases very 
similar. A single combined fit, however, seems somewhat 
advantageous regarding the determination of $\chi_{\rm t}$,
see Subsection~\ref{SEC588}.

\begin{figure}[h!]
\begin{center}
\vspace*{-2mm}
\includegraphics[width=0.7\textwidth]{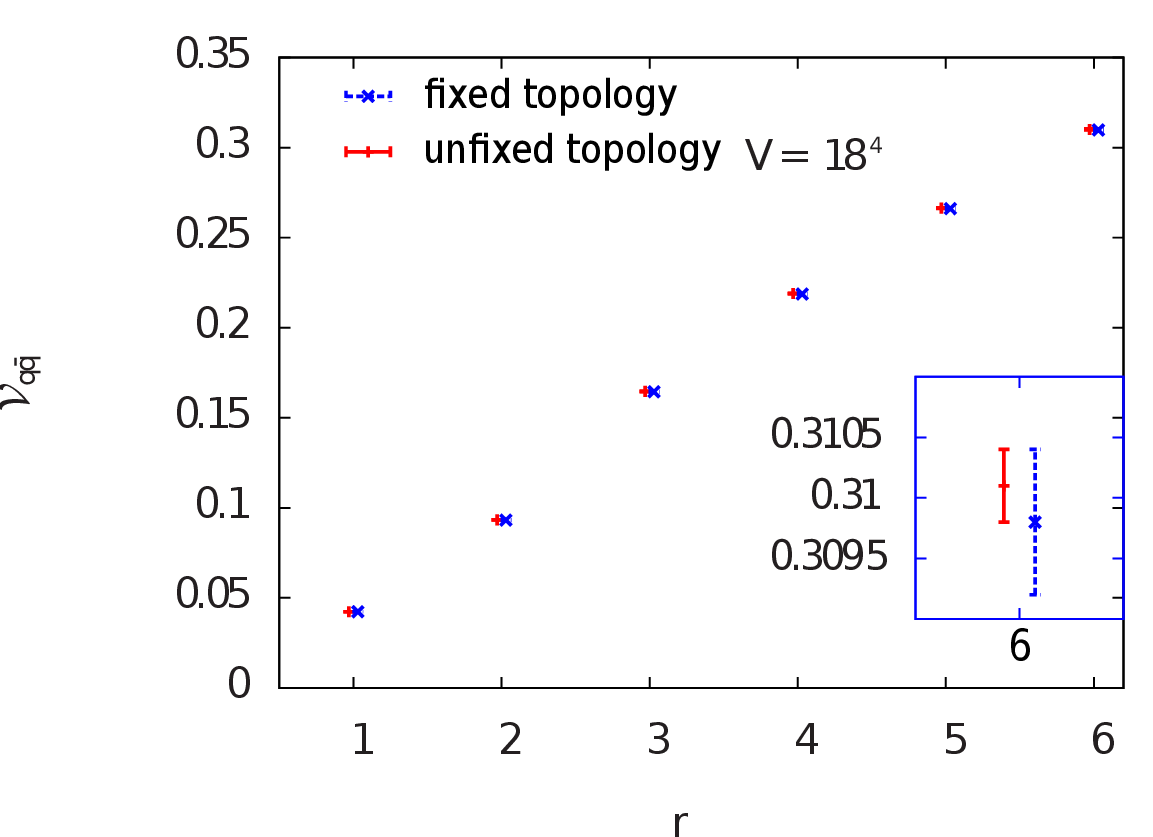}
\vspace*{-2mm}
\caption{\label{FIG003}Comparison of static potential
obtained from fixed topology Wilson loops, in the volumes 
$V = 14^4 , \, 15^4 , \, 16^4 , \, 18^4$, with
$1 / (\chi_{\rm t} V) , \, |Q|/(\chi_{\rm t} V) < 1$, using expansion 
(\ref{EQN004}) with one combined fit, and directly measured at 
$V = 18^4$. (Since unfixed and fixed topology results 
coincide within the errors, they are shifted horizontally
for better visibility.)}
\end{center}
\vspace*{-5mm}
\end{figure}

Figure~\ref{FIG003} compares the static potential obtained 
from fixed topology Wilson loops,
and computed without topology fixing at $V = 18^4$. 
As reflected by Table~\ref{TAB001} there is excellent agreement 
within the errors.

\begin{figure}[htb]
\begin{center}
\vspace*{-2mm}
\includegraphics[width=0.90\textwidth]{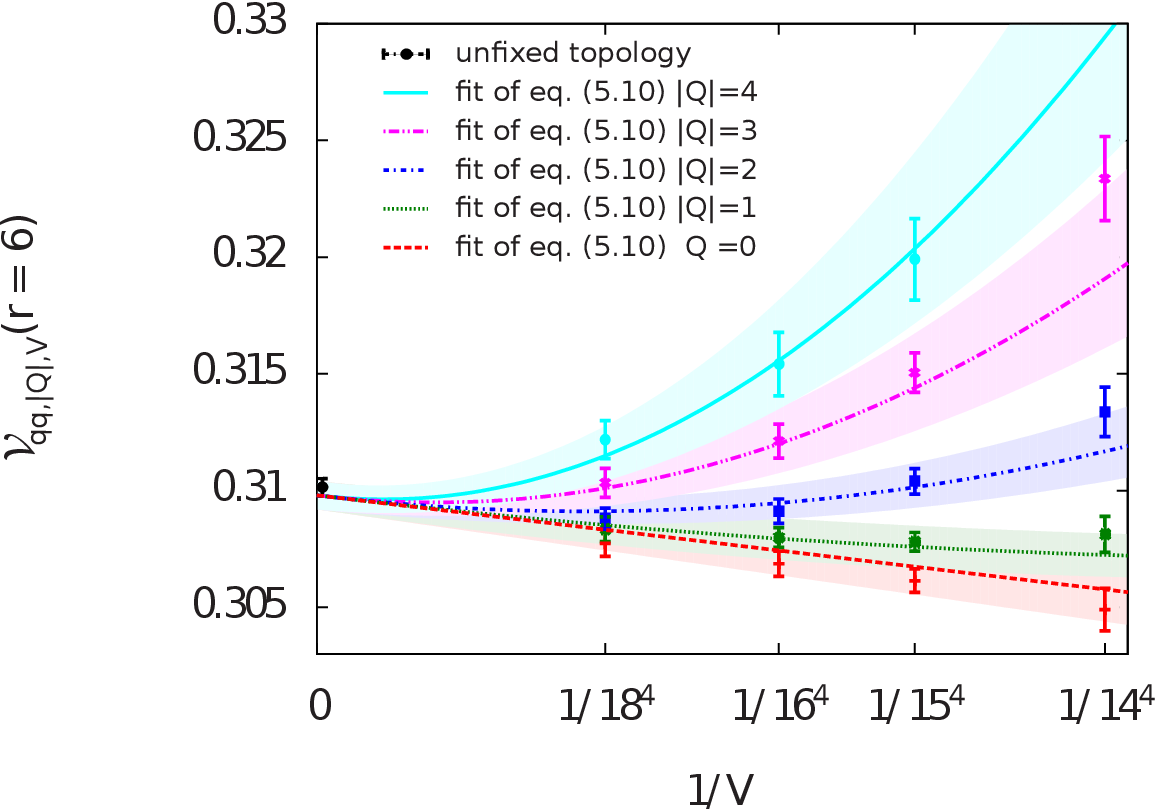}
\vspace*{-2mm}
\caption{\label{FIG690}The fixed topology static potential 
$\mathcal{V}_{q \bar{q},Q,V}(r=6)$ for $|Q| = 0 \ldots 4$, 
as a function of $1/V$, and the curves corresponding to
approximation (\ref{EQN491}).}
\vspace*{-5mm}
\end{center}
\end{figure}

The expansion (\ref{EQN003}) of fixed topology Wilson loop 
averages $\la W_{V}(r,t) \ra_{|Q|}$ is a decaying exponential 
in $t$. This suggests defining a static potential at fixed 
topological charge $|Q|$ and volume $V$,
\begin{eqnarray}
\label{EQN587} \mathcal{V}_{q \bar{q},|Q|,V}(r) = 
-\frac{d}{dt} \ln\Big(\Big\la W_{V}(r,t) \Big\ra_{|Q|} \Big)
\end{eqnarray}
for some value of $t$, where formula (\ref{EQN003}) is a rather 
precise approximation. Within statistical errors 
$\mathcal{V}_{q \bar{q},|Q|,V}(r)$ is independent of $t$ for 
$t_\textrm{min} \leq t \leq t_\textrm{max}$. Therefore, we 
determine $\mathcal{V}_{q \bar{q},|Q|,V}(r)$ by a $\chi^2$ 
minimizing fit of a constant to the right-hand-side of 
eq.\ (\ref{EQN587}), with the derivative replaced by a finite 
difference (this is the common definition of an effective 
mass) in the interval $t_\textrm{min} \leq t \leq t_\textrm{max}$. 
For $|Q| = 0 \ldots 4$ and $V = 14^4, 15^4, 16^4, 18^4$, the 
values for $\mathcal{V}_{q \bar{q},|Q|,V}(r=6)$ are 
plotted in Figure~\ref{FIG690}. As already shown in 
Figure~\ref{FIG002}, there is a strong dependence 
on the topological sector, which becomes increasingly prominent 
for smaller volumes. From expansion (\ref{EQN003}) the fixed 
topology static potential is expected to behave as
\begin{eqnarray} \label{EQN491} 
\mathcal{V}_{q \bar{q},|Q|,V}(r) \approx {\mathcal V}_{q \bar{q}}(r) 
+ \frac{1}{2} {\mathcal V}''_{q \bar{q}}(r) \frac{1}{V \chi_{\rm t}} 
\bigg(1 - \frac{Q^2}{V \chi_{\rm t}}\bigg) \ .
\end{eqnarray}
The corresponding curves for $|Q| = 0 \ldots 4$, with 
parameters ${\mathcal V}_{q \bar{q}}(r=6)$, 
${\mathcal V}''_{q \bar{q}}(r=6)$ and $\chi_{\rm t}$ determined by 
the previously discussed fits
($V = 14^4 \dots 18^4$, $1 /(\chi_{\rm t} V)$,$\, |Q|/(\chi_{\rm t} V) < 1$, 
expansion (\ref{EQN003}) and a single combined fit),
are also shown in Figure~\ref{FIG690}. One clearly sees 
that approximation (\ref{EQN491}) nicely describes the 
numerical results for $\mathcal{V}_{q \bar{q},|Q|,V}(r=6)$.

We conclude that one can obtain a correct and accurate physical 
static potential from Wilson loops separately computed in different 
topological sectors. The errors are 
larger by factors $\approx 2 \ldots 5$ (cf.\ Table~\ref{TAB001}) 
for a fixed topology computation using four ensembles,
compared to a corresponding direct computation using 
a single ensemble ($V = 18^4$).

\subsubsection{\label{SEC588}The topological susceptibility}

In Table~\ref{TAB003} we present results for the topological 
susceptibility extracted from fixed topology Wilson loops 
$\la W_{V}(r,t) \ra_{|Q|}$. Again we use the $1/V$ expansion 
(\ref{EQN003}) or (\ref{EQN004}), the constraints 
$1 / (\chi_{\rm t} V) , \, |Q|/(\chi_{\rm t} V) < 1$ or $ < 0.5$, 
and either a single combined fit to all considered separations 
$r = 1 \ldots 6$, or six separate fits, one for each $r$. 
The latter yields six different results for $\chi_{\rm t}$.

\begin{table}[htb]
\hspace*{-8mm}
\begin{tabular}{|c|c|c|c|c|c|c|}
\hline
method & $\mathcal{V}_{q \bar{q}}(1)$ & $\mathcal{V}_{q \bar{q}}(2)$ 
& $\mathcal{V}_{q \bar{q}}(3)$ & $\mathcal{V}_{q \bar{q}}(4)$ 
& $\mathcal{V}_{q \bar{q}}(5)$ & $\mathcal{V}_{q \bar{q}}(6)$\tabularnewline
\hline
\hline
 \multicolumn{7}{|c|}{fixed topology,$\quad$ $V \in 
\{ 14^4 , \, 15^4 , \, 16^4 , \, 18^4 \}$,$\quad$ 
$1 / (\chi_{\rm t} V) , \, |Q|/(\chi_{\rm t} V) < 1$}\tabularnewline
\hline
(\ref{EQN003})c & \multicolumn{6}{c|}{8.8(0.5)}\tabularnewline
\hline
(\ref{EQN003})s & 8.8(0.5) & 8.7(0.6) & 8.6(0.7) & 8.6(0.9) 
& 8.8(1.0) & 8.9(1.2)\tabularnewline
\hline 
(\ref{EQN004})c & \multicolumn{6}{c|}{7.1(0.6)}\tabularnewline
\hline
(\ref{EQN004})s & 8.6(0.5) & 8.2(0.7) & 7.7(0.8) & 7.3(0.9) 
& 7.0(1.0) & 6.7(1.1)\tabularnewline
\hline 
\hline
 \multicolumn{7}{|c|}{fixed topology,$\quad$ $V \in 
\{ 14^4 \, , \, 15^4 \, , \, 16^4 \, , \, 18^4 \}$,$\quad$ 
$1 / (\chi_{\rm t} V) , \, |Q|/(\chi_{\rm t} V) < 0.5$}\tabularnewline
\hline
(\ref{EQN003})c & \multicolumn{6}{c|}{11.8(5.9)}\tabularnewline
\hline
(\ref{EQN003})s & 10.0(14.0) & 20.7(44.3) & 11.1(8.2)\phantom{0} 
& 11.8(16.0) & 12.8(8.7)\phantom{0} & 15.4(52.1)\tabularnewline
\hline 
(\ref{EQN004})c & \multicolumn{6}{c|}{11.9(5.4)}\tabularnewline
\hline 
(\ref{EQN004})s & 10.2(21.8) & 10.7(12.5) & 11.3(8.7)\phantom{0} 
& 11.8(5.8)\phantom{0} & 13.0(9.7)\phantom{0} & 14.6(12.2)\tabularnewline
\hline 
\end{tabular}
\hspace*{1mm}
\caption{\label{TAB003}Results for the topological susceptibility 
$\chi_{\rm t} \times 10^5$ from fixed topology computations of the 
static potential ${\mathcal V}_{q \bar{q}}(r)$ for various separations 
$r = 1 \ldots 6$. In the column ``method'' the equation number 
of the expansion is listed, ``c'' denotes a single combined fit 
for all separations $r=1 \ldots 6$, and ``s'' denotes a separate 
fit for each separation. The reference value from a direct computation 
is $\chi_{\rm t} \times 10^5 = (7.0 \pm 0.9)$ \cite{Forcrand97}.}
\end{table}

Not all of the extracted $\chi_{\rm t}$ values perfectly agree 
with each other or with the result $\chi_{\rm t} = (7.0 \pm 0.9) 
\times 10^{-5}$ from Ref.\ \cite{Forcrand97}, which we 
take as a reference. Using the weak constraint 
$1 / (\chi_{\rm t} V) ,\, |Q|/(\chi_{\rm t} V) < 1$ there seems to be 
a slight tension in form of $\approx 2 \sigma$ discrepancies, when 
performing fits with formula (\ref{EQN003}). The extended expansion 
(\ref{EQN004}) gives somewhat better results: no tension shows up, 
and most results agree with the reference value within $\sigma$.

One might hope for further improvement by using the stronger 
constraint $1 / (\chi_{\rm t} V) , \, |Q|/(\chi_{\rm t} V) < 0.5$,
since then formulae (\ref{EQN003}) and (\ref{EQN004})
are more accurate.
Indeed this leads to consistency with the reference value,
but in most cases the errors are very large, of the order 
of 100\,\% or even more. For this strong constraint 
the available ${\mathcal V}_{q \bar{q},|Q|}$-data are not 
sufficient to extract a useful result for $\chi_{\rm t}$. 
Note that here the error for one combined fit is significantly 
smaller than those for the separate fits.

We conclude that --- in principle --- one can extract the topological 
susceptibility in Yang-Mills theory from the static potential 
at fixed topology using formulae like (\ref{EQN003}) or 
(\ref{EQN004}). In practice, however, one needs precise data
in several large volumes. Only when a variation of the input data
({\it e.g.}\ by using different bounds with respect to 
$1 / (\chi_{\rm t} V) ,\, |Q|/(\chi_{\rm t} V)$) leads to precise 
and stable $\chi_{\rm t}$ values, should one consider the result 
trustworthy. The data used in this work are not sufficient to 
achieve this standard. As we mentioned before, more promising methods
to determine $\chi_{\rm t}$ from simulations at fixed topology using 
the same lattice setup have recently been explored
\cite{topdense,AFHOvari,LSD,slab}.

\section{Results in the Schwinger model}

\subsection{\label{SEC101}Simulation setup}

We finally proceed to the Schwinger model --- or 2d Quantum 
Electrodynamics --- as a test model with dynamical fermions.
This model has the continuum Lagrangian
\begin{equation}
\mathcal{L}_{\rm cont}(\psi,\bar{\psi},A) = 
\sum_{{\rm f}=1}^{N_{\rm f}} \bar{\psi}^{\rm (f)}
\Big(\gamma_{\mu}(\partial_{\mu}+\ri g_{\rm cont} A_{\mu}) +
m^{\rm (f)}\Big)\psi^{\rm (f)} +\frac{1}{4}F_{\mu\nu}F_{\mu\nu} \ ,
\end{equation}
where $N_{\rm f}$ is the number of fermion flavors. It
is a widely used toy model, which shares important features
with QCD. In particular the U(1) gauge theory in two (spacetime)
dimensions allows for topologically non-trivial gauge configurations, 
similar to instantons in 4d Yang-Mills theories and in QCD. 
The topological charge is given by 
\begin{equation}
Q[A] = \frac{1}{\pi}\int d^{2}x\,\epsilon_{\mu\nu} F_{\mu\nu} \ .
\end{equation}
Moreover, for $N_{\rm f}=2$ the low lying energy eigenstates contain
a light iso-triplet composed of quasi Nambu-Goldstone bosons,
which we are going to denote as ``pions''. 
This model also exhibits fermion confinement.

We simulated the Schwinger model on periodic lattices
of volume $V = L \times L$ (as before we use lattice units),
with $N_{\rm f}=2$ mass degenerate flavors. They are represented by
Wilson fermions, and we use the standard plaquette gauge action
(see {\it e.g.}\ Ref.\ \cite{Rothe}).

One can approach the continuum limit by increasing $L$, while
keeping the terms $g L$ and $M_{\pi} L$ fixed, where $M_{\pi}$ 
denotes the pion mass.\footnote{In physical units, $g$ has the 
dimension of a mass, so these products are both dimensionless.
This also introduces a dimensional lattice spacing $a \propto g$.} 
This requires decreasing both $g$ and $M_{\pi}$ 
proportional to $1/L$ (for the latter the fermion mass has to 
be adjusted). It is also common to refer to $\beta = 1/g^{2}$,
in analogy to the previous sections.

As in Sections 3 and 4, we employ a geometric definition of 
the topological charge on the lattice \cite{Luescher82},
\begin{eqnarray}  \label{QUP}
Q[U] = \frac{1}{2\pi}\sum_{P} \phi(P) \ ,
\end{eqnarray}
where $\sum_{P}$ denotes the sum over all 
plaquettes $P = e^{\ri \phi(P)}$, $-\pi < \phi(P) \leq \pi$.
With this definition, $Q \in \mathbb{Z}$ holds for any stochastic 
gauge configuration.

\begin{figure}[htb]
\vspace*{-3mm}
\begin{centering}
\includegraphics[scale=0.32]{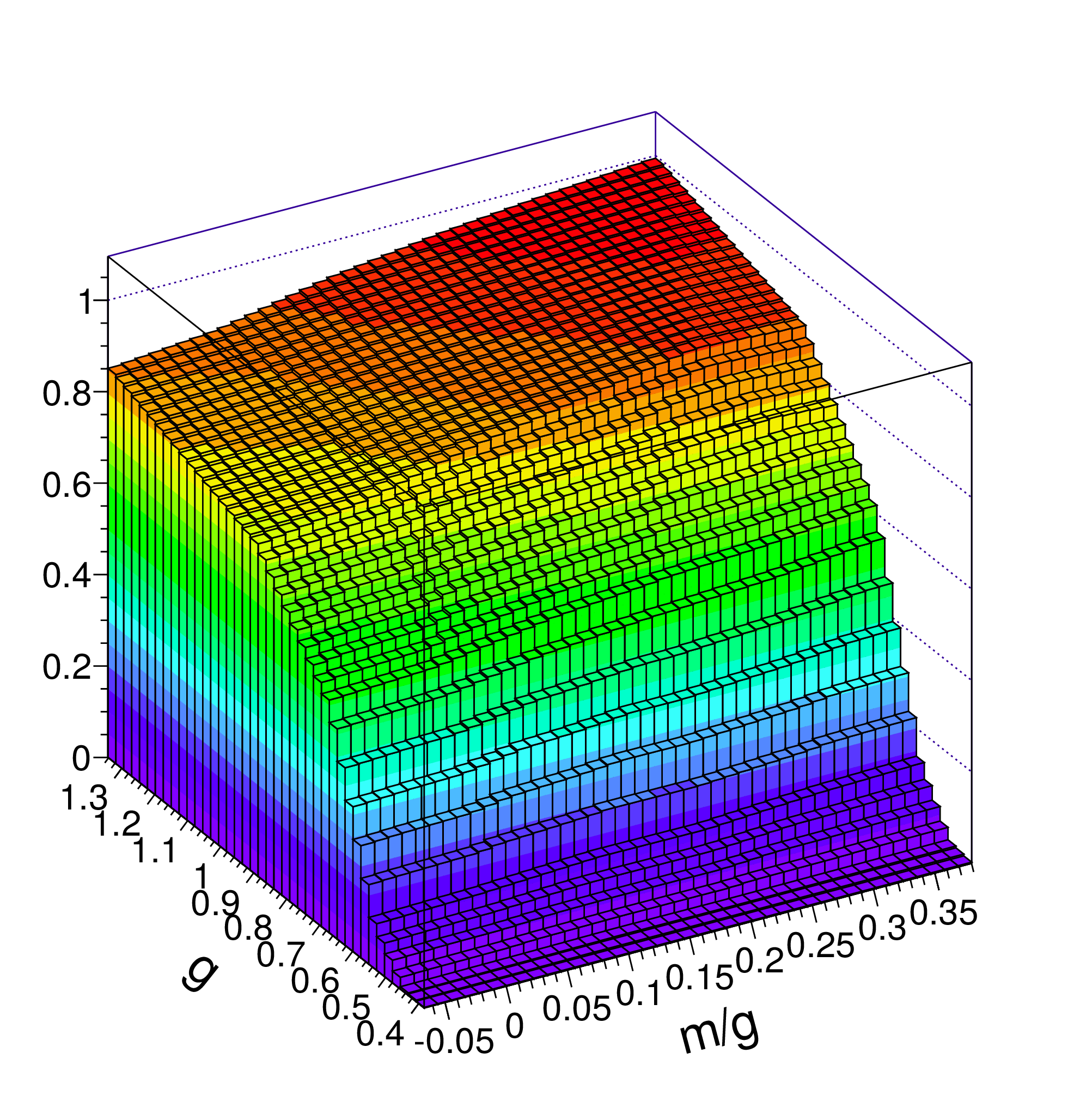}
\par\end{centering}
\centering{}\caption{\label{FIG004xxx}The transition probability 
to a different topological sector per HMC trajectory as 
a function of $g = 1/\sqrt{\beta}$ (varying the lattice 
spacing in physical units, $a \propto g$) and $m/ g$ 
(varying the bare fermion mass in physical units) at $g L = 24/\sqrt{5}$ 
(fixed dimensional volume and coupling constant).}
\end{figure}

We performed simulations at various values of $\beta$, $m$
and $L$ using the HMC algorithm of Ref.\ \cite{HMC_Urbach}, 
with multiple timescale integration and mass preconditioning 
\cite{Hasenbusch}. We started with rather short simulations 
($\approx 50 \, 000 \ldots 100 \, 000$ HMC trajectories)
on small lattices ($L = 8 \ldots 28$),
to investigate the transition probability between 
topological sectors per HMC trajectory. This probability is plotted 
in Figure~\ref{FIG004xxx}, as a function of $g = 1/\sqrt{\beta}$
and $m / g$, while $g L = 24/\sqrt{5}$ is kept constant. 
(The ratio $m/g$ is proportional to the bare fermion mass in 
physical units.)
As expected, topological transitions are frequent at large 
couplings $g$ (coarse lattices), whereas at weak coupling
(fine lattices) topology freezing is observed. 
Such a freezing is also observed in QCD,
which is the main motivation
of this work. We see that the dependence of the transition probability 
on the ratio $m/ g$, and therefore on the dimensional bare fermion mass, 
is rather weak.

Similarly to the previous two sections we now explore the possibility
of extracting physical energy levels (the ``hadron'' masses in the 
Schwinger model) from simulations at fixed topology. To obtain
such results with small statistical errors, we focused on
a single coupling and a single ``quark'' mass,
\be  \label{paramschwing}
\beta=4 \ , \quad m=0.1 \ ,
\ee
and we performed long simulations ($\approx 500 \, 000$ HMC trajectories) 
for volumes $V = L \times L$, with $L = 40,\, 44\, ,48\, ,52\, ,56\, ,60$. 

\subsection{\label{SEC488}Computation of observables}

We determine the topological charge $Q[U]$ for each gauge configuration
$U$ according to definition (\ref{QUP}).
(To measure observables at fixed topological charge $\nu$, 
we only use the configurations with $Q[U]=\nu$.)

The hadron masses that we investigate are the static potential
$\mathcal{V}_{\bar{q}q}(r)$, which has been discussed before in
Yang-Mills theory (Subsection~\ref{SEC010}), and the pion mass 
$M_{\pi}$. A suitable pion creation operator reads
\begin{eqnarray}
O_{\pi} = \sum_{x}\bar{\psi}^{(u)}_{x} \gamma_{3} \psi^{(d)}_{x} \ ,
\end{eqnarray}
where $u$ and $d$ label the two (degenerate) fermion flavors.\footnote{For 
an introduction about the construction of hadron creation 
operators, see {\it e.g.}\ Ref.\ \cite{Weber}.}
For the static potential we use again
\begin{equation}
O_{q\bar{q}}=\bar{q}(r_{1})U(r_{1};
r_{2})q(r_{2}) \ , \quad r=| r_{1} - r_{2} | \ .
\end{equation}
Also here $\bar{q}$ and $q$ represent spinless
static fermions and $U(r_{1}; r_{2})$ denotes the
product of spatial links connecting the lattice sites $r_{1}$ and $r_{2}$
on a given time slice. 
Since there is only one spatial dimension, we do not apply
any gauge link smearing.

\subsection{Numerical results}

\subsubsection{The pion mass and the static potential}

Similar to eq.\ (\ref{EQN587}) one can define a pion mass at 
fixed topological charge $|Q|$ and volume $V$ by
\begin{equation}  \label{EQN587_} 
M_{\pi,|Q|,V} = -\frac{d}{dt} 
\ln\Big(\Big\langle O_\pi^\dagger(t) O_\pi(0) \Big\rangle\Big)
\end{equation}
for some value of $t$, where approximation (\ref{EQN003}) is
quite precise. Within statistical errors, $M_{\pi,|Q|,V}$ is 
independent of $t$ for large $t$. Therefore, 
we determine $M_{\pi,|Q|,V}$ by a $\chi^2$ minimizing fit of a 
constant to the right-hand-side of eq.\ (\ref{EQN587_}) (with the 
derivative replaced by a finite difference).

\begin{figure}
\begin{centering}
\includegraphics[scale=1.0]{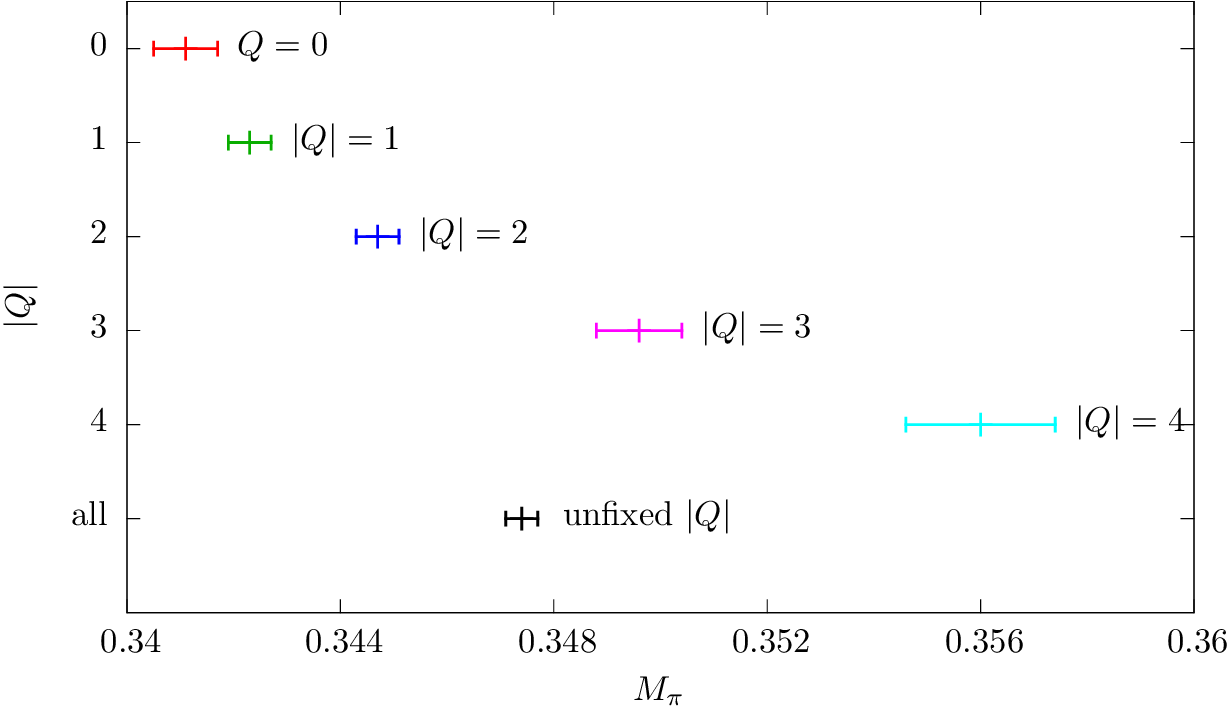}
\caption{\label{fig:Pion-mass-discrepnecy}The pion masses 
$M_{\pi ,|Q|}$ in distinct topological sectors $|Q| = 0 \ldots 4$, 
and $M_{\pi}$ obtained in all sectors, in the volume $V =40^2$.}
\end{centering}
\end{figure}

Figure~\ref{fig:Pion-mass-discrepnecy} shows that pion masses 
obtained at fixed topology in different topological sectors, 
$M_{\pi, |Q|}$, differ significantly at $V = 40^2$. 
For example $M_{\pi,0}$ and $M_{\pi,3}$
differ by more than $6 \sigma$.
The physically meaningful value measured in all sectors, $M_{\pi}$,
also deviates {\it e.g.}\ from $M_{\pi ,0}$ by more than $7 \sigma$. 
Figure~\ref{fig:Pion-mass-discrepnecy} demonstrates also here
the necessity to analytically assemble fixed topology results,
when the Monte Carlo algorithm is unable to generate
frequent changes in $Q$. 

To determine the pion mass and the static potential from correlation 
functions evaluated in single topological sectors, 
$M_{\pi , |Q|}$ and ${\cal V}_{q \bar q, |Q|}$,
we follow the lines of Section 5. 
We perform least-square fits using expansion 
(\ref{EQN003}) or (\ref{EQN004}) of the correlation functions.
We choose a suitable fitting range $t_\textrm{min} \ldots t_\textrm{max}$,
which typically leads to
$\chi^2/\textrm{d.o.f.} \ltapprox 1$. 
The stability of the resulting $M_{\pi , |Q|}$ and ${\cal V}_{q \bar q, |Q|}$
has been checked by varying $t_\textrm{min}$ and $t_\textrm{max}$
by $\pm 1$. The $t$ ranges used for the determination of the pion 
mass are listed in Table~\ref{tab:Temporal-fitting-ranges-schwinger}.

\begin{table}
\centering
\begin{tabular}{|c|c|c|c|c|}
\hline 
$V$ & $t_\textrm{min}$ & $t_\textrm{max}$ & maximum $|Q|$ 
for  & maximum $|Q|$ for \tabularnewline
 &  &  & $1/(\chi_{\rm t}V)$, $|Q|/(\chi_{\rm t}V) < 1$  
& $1/(\chi_{\rm t}V)$, $|Q|/(\chi_{\rm t}V) < 0.5$ \tabularnewline
\hline 
$40^{2}$ & $12$ & $16$ & $7$ & $3$\tabularnewline
$44^{2}$ & $12$ & $18$ & $9$ & $4$\tabularnewline
$48^{2}$ & $12$ & $20$ & $11$ & $5$\tabularnewline
$52^{2}$ & $12$ & $22$ & $13$ & $6$\tabularnewline
$56^{2}$ & $12$ & $24$ & $15$ & $7$\tabularnewline
$60^{2}$ & $12$ & $24$ & $17$ & $8$\tabularnewline
\hline 
\end{tabular}
\caption{\label{tab:Temporal-fitting-ranges-schwinger}Temporal 
fitting ranges $t_\textrm{min} \ldots t_\textrm{max}$ and maximum 
topological charges $|Q|$ for the volumes $V$ under consideration.}
\end{table}

We perform fits in three different ways: (``c'')~a 
single combined fit to all five observables ($M_\pi$, 
$\mathcal{V}_{q \bar{q}}(r = 1)$, $\mathcal{V}_{q \bar{q}}(r = 2)$, 
$\mathcal{V}_{q \bar{q}}(r = 3)$, $\mathcal{V}_{q \bar{q}}(r = 4)$); 
(``c$\mathcal{V}$'')~a single combined fit to the four static 
potential observables; (``s'')~five separate fits, one to each 
of the five observables. The results are collected in 
Table~\ref{tab:Results-schwinger}, along with reference values 
obtained in all sectors at $V = 60^2$.\footnote{In the continuum 
2-flavor Schwinger model, the pion mass is predicted as \cite{Smilga} 
$M_{\pi,{\rm cont}} = 2.008 \dots \times (m_{\rm cont}^{2} g_{\rm cont} )^{1/3}$. 
Remarkably, there is almost perfect agreement with our 
result for $M_{\pi}$, if we insert the bare fermion mass and $\beta$
given in eq.\ (\ref{paramschwing}), which yields 
$M_{\pi} \simeq 0.343$.\label{fnschwing}}

The conclusions are essentially the same as for Yang-Mills theory 
discussed in Section 5. Results extracted indirectly, 
from simulations at fixed topology, are in agreement with those 
obtained directly. The magnitude of 
the errors is the same for the two expansions 
(\ref{EQN003}) and (\ref{EQN004}), and for the fitting methods ``c'', 
``c$\mathcal{V}$'' and ``s''. They are, however, larger by factors 
of $\approx 2$ when we use the stringent constraint 
$1/(\chi_{\rm t} V) , \, |Q|/(\chi_{\rm t} V) < 0.5$, since 
less input data are involved compared to 
$1/(\chi_{\rm t} V) ,\, |Q|/(\chi_{\rm t} V) < 1$. 
The fits all yield uncorrelated $\chi^{2} / \textrm{d.o.f.} 
\ltapprox 1$, indicating that the fixed topology results are well 
described by both formulae (\ref{EQN003}) and (\ref{EQN004}).

\begin{table}
\hspace*{-2.5mm}
\begin{tabular}{|c|c|c|c|c|c|}
\hline 
method & $M_\pi$ & $\mathcal{V}_{q \bar{q}}(1)$ & 
$\mathcal{V}_{q \bar{q}}(2)$ & $\mathcal{V}_{q \bar{q}}(3)$ 
& $\mathcal{V}_{q \bar{q}}(4)$\tabularnewline
\hline 
\hline 
\multicolumn{6}{|c|}{all sectors,$\quad$ $V = 60^{2}$}\tabularnewline
\hline 
 & 0.3474(3) & 0.1296(2) & 0.2382(5) & 0.3288(7) & 0.4045(10)\tabularnewline
\hline 
\hline 
\multicolumn{6}{|c|}{fixed topology,~ 
$V \in\{40^{2},44^{2},48^{2},52^{2},56^{2},60^{2}\}$,~  
$1/(\chi_{\rm t}V)$, $|Q|/(\chi_{\rm t}V) < 1$ }\tabularnewline
\hline 
(\ref{EQN003})c & 0.3466(16) & 0.1293(19) & 0.2370(23) 
& 0.3261(29) & 0.4022(62)\tabularnewline
(\ref{EQN003})c$\mathcal{V}$ &  & 0.1295(10) & 0.2372(12) 
& 0.3386(15) & 0.4052(16)\tabularnewline
(\ref{EQN003})s & 0.3477(8) ~~\!\!\! &  0.1285(7) ~~\!\!\! 
& 0.2371(9) ~~\!\!\! & 0.3282(12) & 0.4050(16)\tabularnewline
\hline 
(\ref{EQN004})c & 0.3467(10) & 0.1293(6) ~~\!\!\! & 0.2377(9) ~~\!\!\!
& 0.3321(32) & 0.4059(69)\tabularnewline
(\ref{EQN004})c$\mathcal{V}$ &  & 0.1295(5) ~~\!\!\! & 0.2379(11) 
& 0.3392(14) & 0.4049(16)\tabularnewline
(\ref{EQN004})s & 0.3477(9)  ~~\!\!\! & 0.1294(5) ~~\!\!\! & 
0.2374(6) ~~\!\!\! & 0.3288(12) & 0.4040(15)\tabularnewline
\hline 
\hline 
\multicolumn{6}{|c|}{fixed topology,~ 
$V \in\{40^{2},44^{2},48^{2},52^{2},56^{2},60^{2}\}$,~
$1/(\chi_{\rm t}V)$, $|Q|/(\chi_{\rm t}V) < 0.5$ }\tabularnewline
\hline 
(\ref{EQN003})c & 0.3454(32) & 0.1284(27) & 0.2364(28) 
& 0.3311(50) & 0.4049(80)\tabularnewline
(\ref{EQN003})c$\mathcal{V}$ &  & 0.1282(12) & 0.2370(16) 
& 0.3312(35) & 0.4175(82)\tabularnewline
(\ref{EQN003})s & 0.3478(32) & 0.1292(12) & 0.2377(21) 
& 0.3275(61) & 0.4027(91)\tabularnewline
\hline 
(\ref{EQN004})c & 0.3455(32) & 0.1285(16) & 0.2365(19) 
& 0.3310(49) & 0.4048(78)\tabularnewline
(\ref{EQN004})c$\mathcal{V}$ &  & 0.1287(9) ~~\!\!\! & 0.2371(23) 
& 0.3312(36) & 0.4073(83)\tabularnewline
(\ref{EQN004})s & 0.3482(35) & 0.1291(11) & 0.2376(13) 
& 0.3290(22) & 0.4036(55)\tabularnewline
\hline 
\end{tabular}
\caption{\label{tab:Results-schwinger}Results for the pion 
mass $M_{\pi}$ and the static potential ${\mathcal{V}}_{q\bar{q}}(r)$ 
at separations $r=1,2,3,4$, with and without
topology fixing. In the column ``method'' the equation 
number of the expansion is listed, ``c'' denotes one 
combined fit to all five observables, ``c$\mathcal{V}$'' 
means one combined fit to the four static potential 
observables, and ``s'' indicates separate fits for each of 
the five observables.}
\end{table}

\begin{figure}[htb]
\centering
\includegraphics{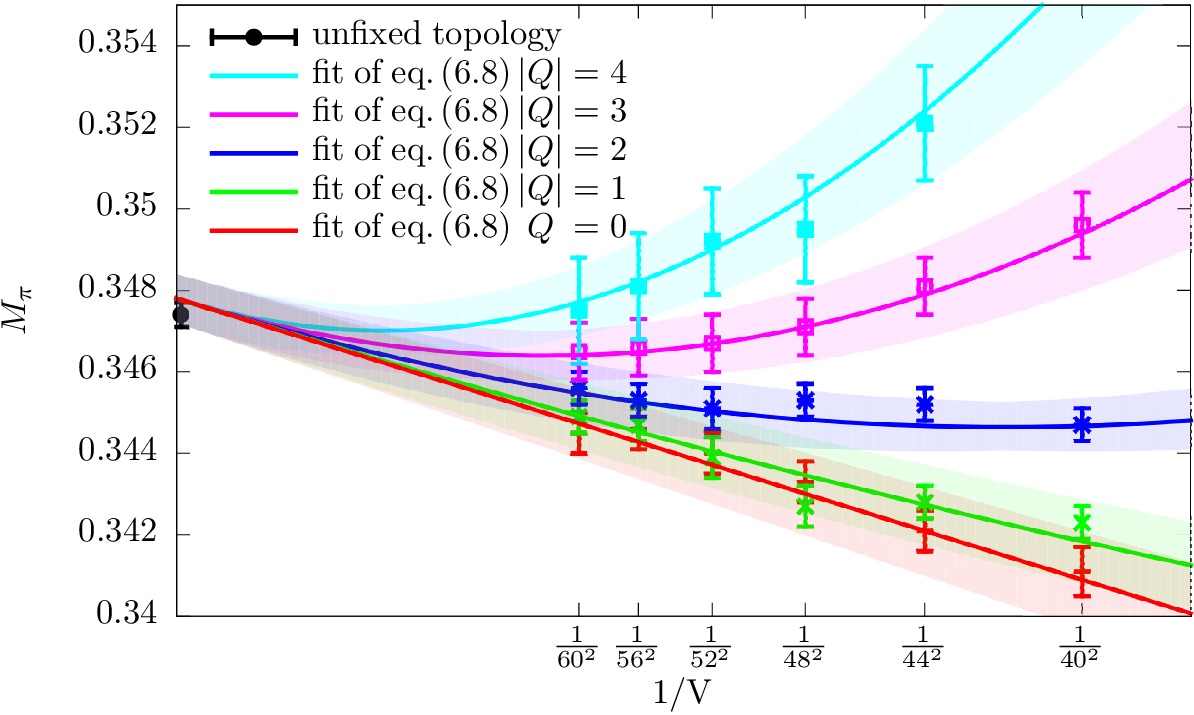}
\caption{\label{fig:pion_BCNW}The fixed topology pion mass 
$M_{\pi,|Q|,V}$ for $|Q|=0 \dots 4$, as a function of $1/V$, 
and the curves corresponding to formula (\ref{EQN491_}).}
\end{figure}

For $|Q| = 0 \ldots 4$ and $V = 40^2 \dots 60^2$, 
the $M_{\pi,|Q|,V}$ values are plotted in 
Figure~\ref{fig:pion_BCNW}. Again we observe a strong dependence 
on the topological sector, in particular in 
small volumes. From the expansion (\ref{EQN003}), $M_{\pi,|Q|,V}$ 
is expected to behave as approximation (\ref{approx2}),
\be
\label{EQN491_}M_{\pi,Q,V} = M_{\pi} + 
\frac{c}{V \chi_{\rm t}} \bigg(1 - \frac{Q^2}{V \chi_{\rm t}}
\bigg) \ , \quad c = \frac{1}{2} M_{\pi}''(\theta )_{\pi}|_{\theta =0} \ .
\ee
The corresponding curves for $|Q| = 0 \ldots 4$ with 
parameters $M_{\pi}$, $M''_{\pi}$ and $\chi_{\rm t}$, determined by 
the previously discussed fit ``(\ref{EQN003})s'', are also 
shown in Figure~\ref{fig:pion_BCNW}. One can clearly see 
that approximation (\ref{EQN491_}) nicely captures the
lattice results for $M_{\pi,|Q|,V}$.

We conclude, similar to our study in Yang-Mills theory,
that it is possible to extract correct and accurate values for 
the pion mass and the static potential from correlation functions 
computed in a number of fixed topological sectors and volumes. 
The errors are somewhat larger 
than for direct computation, in our case
by factors of $\approx 2 \ldots 7$. This is partly 
due to the smaller amount of gauge configurations of the fixed 
$Q$ ensembles at different $V$, and partly due to the extrapolation 
to infinite volume.

\subsubsection{The topological susceptibility}

Table~\ref{tab:Results-schwinger-sus} presents results 
for the topological susceptibility extracted from our data for
$M_{\pi, |Q|}$ and ${\cal V}_{q \bar q, |Q|}$.
These values for $\chi_{\rm t}$ are obtained 
from the same fits, which lead to the
results in Table~\ref{tab:Results-schwinger}. 
The results for $\chi_{\rm t}$ and their interpretation are 
similar to those obtained in Yang-Mills theory. 
We observe a slight tension of $\approx 2 \sigma$ for some 
values, when using expansion (\ref{EQN003}) and the relaxed 
constraint ($1/(\chi_{\rm t} V) , |Q|/(\chi_{\rm t} V) < 1$). This tension 
disappears when we apply the improved expansion (\ref{EQN004}). 
When imposing the strict constraint ($1/\chi_{\rm t} V , \, 
|Q|/\chi_{\rm t} V < 0.5$), we encounter the same problem as in 
Subsection~\ref{SEC588}: all results are in agreement with the 
directly measured $\chi_{\rm t} = \la Q^2 \ra / V$ (at $V=60^{2}$), 
but the errors are very large.\footnote{Ref.\ \cite{DuHo1}
presents results for $\chi_{\rm t}$ in the 2-flavor Schwinger model
with staggered and overlap fermions, with or without link smearing. 
The results at $\beta =4$ and $m=0.1$ (in large volume) are in the 
range $\chi_{\rm t} \simeq 0.044 \dots 0.064$. This agrees with our
value in Table~\ref{tab:Results-schwinger-sus}, which confirms the 
mild renormalization of our bare fermion mass (cf.\ footnote 
\ref{fnschwing}).}

{\small
\begin{table}[htb]
\hspace*{-2.5mm}
\begin{tabular}{|c|c|c|c|c|c|}
\hline 
method & $M_\pi$ & $\mathcal{V}_{q \bar{q}}(1)$ & 
$\mathcal{V}_{q \bar{q}}(2)$ & $\mathcal{V}_{q \bar{q}}(3)$ & 
$\mathcal{V}_{q \bar{q}}(4)$\tabularnewline
\hline 
\hline 
\multicolumn{6}{|c|}{all sectors,
$\quad$ $ V=60^{2} $}\tabularnewline
\hline 
\multicolumn{6}{|c|}{0.0048(1)}\tabularnewline
\hline 
\hline 
\multicolumn{6}{|c|}{fixed topology, \  
$ V \in\{ 40^{2},44^{2},48^{2},52^{2},56^{2},60^{2} \}$, \ 
$1/(\chi_{\rm t}V)$, $|Q|/(\chi_{t}V) < 1$}\tabularnewline
\hline 
(\ref{EQN003})c & \multicolumn{5}{c|}{{\small{0.0038(5)}}}\tabularnewline
\hline 
(\ref{EQN003})c$\mathcal{V}$ &  & \multicolumn{4}{c|}{0.0042(5)}\tabularnewline
\hline 
(\ref{EQN003})s & 0.0041(4) & 0.0038(5) & 0.0036(7) 
& 0.0038(11) & 0.0044(9)\tabularnewline
\hline 
(\ref{EQN004})c & \multicolumn{5}{c|}{{\small{0.0044(4)}}}\tabularnewline
\hline 
(\ref{EQN004})c$\mathcal{V}$ &  & \multicolumn{4}{c|}{0.0042(6)}\tabularnewline
\hline 
(\ref{EQN004})s & 0.0046(5) & 0.0043(4) & 0.0045(7) & 0.0036(12) 
& 0.0038(8)\tabularnewline
\hline 
\hline 
\multicolumn{6}{|c|}{fixed topology, \ 
$V \in \{40^{2},44^{2},48^{2},52^{2},56^{2},60^{2} \}$, \ 
$1/(\chi_{\rm t}V)$, $|Q|/(\chi_{\rm t}V)<0.5$ }\tabularnewline
\hline 
(\ref{EQN003})c & \multicolumn{5}{c|}{{\small{0.0065(35)}}}\tabularnewline
\hline 
(\ref{EQN003})c$\mathcal{V}$ &  & \multicolumn{4}{c|}{0.0017(30)}\tabularnewline
\hline 
(\ref{EQN003})s & 0.0014(38) & 0.0049(32) & 0.0057(31) 
& 0.0037(48) & 0.0032(27)\tabularnewline
\hline 
(\ref{EQN004})c & \multicolumn{5}{c|}{{\small{0.0067(32)}}}\tabularnewline
\hline 
(\ref{EQN004})c$\mathcal{V}$ &  
& \multicolumn{4}{c|}{0.0018(33)}\tabularnewline
\hline 
(\ref{EQN004})s & 0.0017(32) & 0.0043(34) & 0.0022(46) 
& 0.0015(38) & 0.0048(52)\tabularnewline
\hline 
\end{tabular}
\caption{\label{tab:Results-schwinger-sus}Results for the 
topological susceptibility $\chi_{\rm t}$, directly measured
(at $V=60^{2}$), and based on fixed topology computations of $M_{\pi , |Q|}$ 
and $\mathcal{V}_{q\bar{q}, |Q|}(r)$ for separations $r=1,2,3,4$. 
In the column ``method'' the equation number of the expansion 
is listed, ``c'' denotes a single combined fit to all five 
observables, ``c$\mathcal{V}$'' means a single combined 
fit to the four static potential observables, and ``s'' denotes 
a separate fit to each of the five observables.}
\end{table}
}

We infer that a reasonably accurate determination of 
the topological susceptibility from $M_{\pi, |Q|}$ and 
$\mathcal{V}_{q\bar{q}, |Q|}$ requires extremely 
precise input data. The fixed topology ensembles and correlation 
functions of this work are not sufficient 
to extract an accurate and stable value for $\chi_{\rm t}$.

\section{Conclusions}

We have systematically explored the applicability of the
Brower-Chandra\-sekharan-Negele-Wiese (BCNW) method \cite{BCNW}
with lattice data in fixed topological sectors. Our study
encompasses the quantum rotor, the Heisenberg model,
4d SU(2) Yang-Mills theory and the 2-flavor Schwinger 
model. The originally suggested application to the pion mass
has been extended to other observables, like the magnetic
susceptibility and the static quark-antiquark potential.

The primary goal of this method is the determination of a
physical observable if only fixed topology results are
available. Our observations show that this can be achieved
to a good precision with input data from various volumes
and topological sectors, which obey the (rather relaxed)
constraint $1/(\chi_{\rm t} V), \, |Q|/(\chi_{\rm t} V) < 1$.
Hence this method is promising for application in QCD, where
lattice spacings below $a \simeq 0.05 \ {\rm fm}$ are expected
to confine HMC simulations to a single topological sector
over extremely long trajectories.

As a second goal, this method also enables --- in principle --- the
determination of the topological susceptibility $\chi_{\rm t}$.
In our study we obtained the right magnitude also for $\chi_{\rm t}$, 
but the results were usually plagued by large uncertainties. 
For this purpose, {\it i.e.}\ for the measurement of $\chi_{\rm t}$ 
based on fixed topology simulation results, other methods are more 
appropriate, based on the topological charge density correlation
\cite{AFHO,AFHOvari,topdense}, or on an analysis of $\chi_{\rm t}$
in sub-volumes \cite{LSD,slab}.

Regarding the optimal way to apply this method, it seems 
--- for lattice data of typical statistical precision ---
not really helpful to add additional terms of the $1/(\chi_{\rm t} V)$
expansion, beyond the incomplete second order that was suggested in 
Ref.\ \cite{BCNW}. Higher terms were elaborated in Ref.\ \cite{Arthur14},
and they improve the agreement with the fixed topology lattice data, but 
due to the appearance of additional free parameters they hardly improve
the results for the physical observable and for $\chi_{\rm t}$.

A step beyond, which deserves being explored in more detail, is the
inclusion of ordinary finite size effects (not related to topology
fixing) \cite{PolProc}, which even allows for the use of
small volumes (in the terminology of Section 2).

At this point, we recommend the application of the simple formulae
(\ref{approx2}) and (\ref{EQN003}) or (slightly better) (\ref{EQN004}), 
with only three free parameters, for the determination of hadron masses 
in QCD on fine lattices, in particular in the presence of very light quarks.
 \\

\noindent
{\bf Acknowledgements} \ We thank Irais Bautista and Lilian Prado
for their contributions to this project at an early stage, and
Carsten Urbach for providing a simulation code for the Schwinger
model, corresponding advice and helpful discussions.
This work was supported by the {\it Helmholtz International Center 
for FAIR} within the framework of the LOEWE program launched by 
the State of Hesse, and by the Mexican {\it Consejo Nacional de 
Ciencia y Tecnolog\'{\i}a} (CONACYT) through projects CB-2010/155905 and 
CB-2013/222812, as well as DGAPA-UNAM, grant IN107915.
A.D., C.C.\ and M.W.\ acknowledge support by the Emmy Noether Programme 
of the DFG (German Research Foundation), grant WA 3000/1-1, and C.P.H.\ 
was supported through the project {\it Redes Tem\'{a}ticas de 
Colaboraci\'{o}n Acad\'{e}mica 2013,} UCOL-CA-56.
Calculations were performed on the LOEWE-CSC and FUCHS-CSC high-performance 
computer of Frankfurt University, and on the cluster of ICN/UNAM.
We also thank HPC-Hessen, funded by the State Ministry of Higher Education, 
Research and the Arts, for programming advice.

\appendix

\section{Low temperature expansion of the magnetic susceptibility
of the quantum rotor}

Our point of departure is eq.\ (\ref{intchim}) for the magnetic 
susceptibility of the quantum rotor at fixed topology\footnote{Since 
this entire appendix refers to continuous Euclidean time, we skip for
simplicity the subscripts of $\beta_{\rm cont}$ and $L_{\rm cont}$.}
\be 
\chi_{{\rm m},Q} = \int_{0}^{L/2} dt \, e^{( t^{2}/L - t)/(2 \beta)} 
\left[ e^{2 \pi \ri Qt/L} + e^{-2 \pi \ri Qt/L} \right] \ .
\ee
By completing the squares in each term, and defining
\be
z_{0} = \sqrt{\frac{L}{8 \beta}} \ 
\Big( 1 + \frac{4 \pi \ri Q \beta}{L} \Big) 
\ee
we obtain
\bea
\chi_{{\rm m},Q} &=& \sqrt{2 \beta L} \ e^{\frac{2 \pi^{2} \beta Q^{2}}{L}
- \frac{L}{8 \beta}} (-1)^{Q} \Big[ 
\int_{-z_{0}}^{-\pi \ri Q \sqrt{2\beta /L}} dt \, e^{t^{2}} 
+ \int_{-z_{0}^{*}}^{\pi \ri Q \sqrt{2\beta /L}} dt \, e^{t^{2}} \Big] \nn \\
&=& \sqrt{\frac{\pi \beta L}{2}} \ 
e^{\frac{2 \pi^{2} \beta Q^{2}}{L} - \frac{L}{8 \beta}} (-1)^{Q}
\Big[ {\rm erfi}(z_{0}) + {\rm erfi}(z_{0}^{*}) \Big] \nn \\
&=& \sqrt{8 \beta L} \ {\rm Re} \ D(z_{0}) \ .
\eea
We have used two properties of the imaginary error function,
${\rm erfi}(z) = - {\rm erfi}(-z)$ and 
${\rm erfi}(z^{*}) = ({\rm erfi}(z))^{*}$, 
and in the last step we inserted Dawson's function
\be
D(z) = \frac{\sqrt{\pi}}{2} \, e^{-z^{2}} \, {\rm erfi}(z) 
= e^{-z^{2}} \int_{0}^{z} dt \, e^{t^{2}} \ .
\ee
We are interested in the case $L \gg \beta$ where 
$| {\rm arg} \, (\ri z_{0}) | \approx \frac{\pi}{2} < \frac{3 \pi}{4}$,
so we can apply the asymptotic expansion \cite{AbSteg}
\be
D(z_{0}) = \frac{1}{2z_{0}} \sum_{n \geq 0} \frac{(2n-1)!!}{(2 z_{0}^{2})^{n}}
= \frac{1}{2z_{0}} \Big( 1 + \frac{1}{2 z^{2}_{0}} + \frac{3}{4 z^{4}_{0}} 
+ \frac{15}{8 z^{6}_{0}} + {\cal O} (|z_{0}|^{-8}) \Big) \ .
\ee 
If we expand
\be
\chi_{{\rm m},Q} \simeq 4 \beta \ {\rm Re} \ 
\Big[ \frac{1}{1 + 4 \pi \ri Q \beta /L} \  
\Big( 1 + \frac{1}{2 z^{2}_{0}} + \frac{3}{4 z^{4}_{0}} 
+ \frac{15}{8 z^{6}_{0}} \Big) \Big]
\ee
to ${\cal O} ((\beta/L)^{3})$, and insert the infinite volume 
limit $\chi_{\rm m} = 4 \beta$, we arrive at
\be  \label{chiexpeand}
\chi_{{\rm m},Q} = \chi_{\rm m} + \beta \left[ \frac{16 \beta}{L}
+ \frac{64 \beta^{2}}{L^{2}} \Big( 3 - (\pi Q)^{2} \Big)
+ \frac{768 \beta^{3}}{L^{3}} \Big( 5 - 2 (\pi Q)^{2} \Big)
+ {\cal O} \Big( (\frac{\beta}{L})^{4} \Big) \right] \ . \
\ee
By substituting $\chi_{\rm t} = \frac{1}{4 \pi^{2} \beta}$
(which only has exponentially suppressed finite size effects \cite{slab}), 
we confirm to each order given in eq.\ (\ref{chiexpeand})
the expansion that we anticipated in eq.\ (\ref{BCNWchim}); 
it is not altered by the truncation of the Gauss integrals.

\end{document}